\documentclass{JINST}
\title{The LHCb Trigger and its Performance in 2011}
\author{R.~Aaij$^{o}$, 
J.~Albrecht$^{m}$, 
F.~Alessio$^{m}$, 
S.~Amato$^{a}$, 
E.~Aslanides$^{d}$, 
I.~Belyaev$^{i}$, 
M.~van~Beuzekom$^{o}$, 
E.~Bonaccorsi$^{m}$,
R.~Bonnefoy$^{c}$,
L.~Brarda$^{m}$,
O.~Callot$^{e}$,
M.~Cattaneo$^{m}$, 
H.~Chanal$^{c}$,
M.~Chebbi$^{m}$,
X.~Cid~Vidal$^{l}$, 
M.~Clemencic$^{m}$, 
J.~Closier$^{m}$, 
V.~Coco$^{o}$, 
J.~Cogan$^{d}$, 
O.~Deschamps$^{c}$, 
H.~Dijkstra$^{m}$\thanks{Corresponding
author.}, 
C.~Drancourt$^{b}$,
R.~Dzhelyadin$^{j}$, 
M.~Frank$^{m}$, 
M.~Gandelman$^{a}$, 
C.~Gaspar$^{m}$, 
V.V.~Gligorov$^{m}$, 
C.~G\"{o}bel$^{v}$, 
L.A.~Granado~Cardoso$^{m}$, 
Yu.~Guz$^{j}$, 
C.~Haen$^{m}$,
J.~He$^{e}$,
E.~van~Herwijnen$^{m}$, 
W.~Hulsbergen$^{o}$,
R.~Jacobsson$^{m}$, 
B.~Jost$^{m}$, 
T.M.~Karbach$^{m}$, 
U.~Kerzel$^{m}$, 
P.~Koppenburg$^{o}$, 
G.~Krocker$^{f}$, 
C.~Langenbruch$^{m}$,
I. Lax$^{g}$,
R.~Le~Gac$^{d}$, 
R.~Lef\`{e}vre$^{c}$,
J.~Lefran\c{c}ois$^{e}$, 
O.~Leroy$^{d}$, 
L.~Li~Gioi$^{c}$, 
G.~Liu$^{m}$,
F.~Machefert$^{e}$,
I.V.~Machikhiliyan$^{b,i}$,
M.~Magne$^{c}$,
G.~Mancinelli$^{d}$, 
U.~Marconi$^{g}$, 
A.~Mart\'{i}n~S\'{a}nchez$^{e}$, 
M.-N.~Minard$^{b}$, 
S.~Monteil$^{c}$, 
N.~Neufeld$^{m}$, 
V.~Niess$^{c}$, 
S.~Oggero$^{o}$, 
A.~P\'{e}rez-Calero~Yzquierdo$^{k}$, 
P.~Perret$^{c}$, 
M.~Perrin-Terrin$^{d}$, 
B.~Pietrzyk$^{b}$, 
A.~Puig~Navarro$^{n}$, 
G.~Raven$^{p}$, 
P.~Robbe$^{e}$, 
H.~Ruiz$^{k}$, 
M.-H.~Schune$^{e}$, 
R.~Schwemmer$^{m}$, 
J.~Serrano$^{d}$, 
I.~Shapoval$^{q,m}$, 
T.~Skwarnicki$^{u}$, 
B.~Souza~De~Paula$^{a}$, 
P.~Spradlin$^{r}$, 
S.~Stahl$^{f}$, 
V.K.~Subbiah$^{m}$, 
S.~T'Jampens$^{b}$, 
F.~Teubert$^{m}$, 
C.~Thomas$^{t}$, 
M.~Vesterinen$^{m}$, 
M.~Williams$^{s,1}$, 
M.~Witek$^{h}$, 
A.~Zvyagin$^{m}$\\
\llap{$ ^{a}$}Universidade Federal do Rio de Janeiro (UFRJ),\\ Rio de Janeiro, Brazil\\
\llap{$ ^{b}$}LAPP, Universit\'{e} de Savoie, CNRS/IN2P3,\\ Annecy-Le-Vieux, France\\
\llap{$ ^{c}$}Clermont Universit\'{e}, Universit\'{e} Blaise Pascal, CNRS/IN2P3, LPC,\\ Clermont-Ferrand, France\\
\llap{$ ^{d}$}CPPM, Aix-Marseille Universit\'{e}, CNRS/IN2P3,\\ Marseille, France\\
\llap{$ ^{e}$}LAL, Universit\'{e} Paris-Sud, CNRS/IN2P3,\\ Orsay, France\\
\llap{$ ^{f}$}Physikalisches Institut, Ruprecht-Karls-Universit\"{a}t Heidelberg,\\ Heidelberg, Germany\\
\llap{$ ^{g}$}Sezione INFN di Bologna,\\ Bologna, Italy\\
\llap{$ ^{h}$}Henryk Niewodniczanski Institute of Nuclear Physics  Polish Academy of Sciences,\\ Krak\'{o}w, Poland\\
\llap{$ ^{i}$}Institute of Theoretical and Experimental Physics (ITEP),\\ Moscow, Russia\\
\llap{$ ^{j}$}Institute for High Energy Physics (IHEP),\\ Protvino, Russia\\
\llap{$ ^{k}$}Universitat de Barcelona,\\ Barcelona, Spain\\
\llap{$ ^{l}$}Universidad de Santiago de Compostela,\\ Santiago de Compostela, Spain\\
\llap{$ ^{m}$}European Organization for Nuclear Research (CERN),\\ Geneva, Switzerland\\E-mail: Hans.Dijkstra@cern.ch\\
\llap{$ ^{n}$}Ecole Polytechnique F\'{e}d\'{e}rale de Lausanne (EPFL),\\ Lausanne, Switzerland\\
\llap{$ ^{o}$}Nikhef National Institute for Subatomic Physics,\\ Amsterdam, The Netherlands\\
\llap{$ ^{p}$}Nikhef National Institute for Subatomic Physics and VU University Amsterdam,\\ Amsterdam, The Netherlands\\
\llap{$ ^{q}$}NSC Kharkiv Institute of Physics and Technology (NSC KIPT),\\ Kharkiv, Ukraine\\
\llap{$ ^{r}$}School of Physics and Astronomy, University of Glasgow,\\ Glasgow, United Kingdom\\
\llap{$ ^{s}$}Imperial College London,\\ London, United Kingdom\\
\llap{$ ^{t}$}Department of Physics, University of Oxford,\\ Oxford, United Kingdom\\
\llap{$ ^{u}$}Syracuse University,\\ Syracuse, NY, United States\\
\llap{$ ^{v}$}Pontif\'{i}cia Universidade Cat\'{o}lica do Rio de Janeiro (PUC-Rio),\\ Rio de Janeiro, Brazil, associated to $^{a}$\\
\llap{$ ^{1}$}Massachusetts Institute of Technology,\\ Cambridge, MA, United States\\
}
\abstract{
This paper presents the design of the LHCb trigger and its performance on data taken at the LHC in 2011.
A principal goal of LHCb is to perform flavour physics measurements, and the trigger
is designed to distinguish charm and beauty decays from the light quark background.
Using a combination
of lepton identification and measurements of the particles' transverse momenta the trigger
selects particles 
originating from charm and beauty hadrons, which typically fly a finite distance before decaying.
The trigger reduces the roughly $11$\,MHz of bunch-bunch crossings that contain 
at least one inelastic $pp$ interaction to 3\,kHz. This reduction
takes place in two
stages; the first stage is implemented in hardware and the second stage
is a software application that runs on a large computer farm.
A data-driven method is used to evaluate the performance of the trigger on several charm and beauty decay modes.
}


\usepackage{xspace}
\usepackage{subfigure}
\usepackage{epsfig}
\usepackage{ifthen} % for conditional statements
\newboolean{uprightparticles}
\setboolean{uprightparticles}{false} %Set to true to get roman particle symbols
\def\ux85 {UX85\xspace}

\ifthenelse{\boolean{uprightparticles}}%
{

 \def\Ppi         {\ensuremath{\uppi}\xspace}

 \def\Ppsi        {\ensuremath{\uppsi}\xspace}

 \def\PDelta      {\ensuremath{\Delta}\xspace}                 
 \def\PXi      {\ensuremath{\Xi}\xspace}                 
 \def\PLambda      {\ensuremath{\Lambda}\xspace}                 
 \def\PSigma      {\ensuremath{\Sigma}\xspace}                 
 \def\POmega      {\ensuremath{\Omega}\xspace}                 
 \def\PUpsilon      {\ensuremath{\Upsilon}\xspace}                 
 
 \def\PB      {\ensuremath{\mathrm{B}}\xspace}                 
                  
 \def\PD      {\ensuremath{\mathrm{D}}\xspace}

 \def\PJ      {\ensuremath{\mathrm{J}}\xspace}                 
 \def\PK      {\ensuremath{\mathrm{K}}\xspace}

 \def\PW      {\ensuremath{\mathrm{W}}\xspace}

 \def\PZ      {\ensuremath{\mathrm{Z}}\xspace}

 \def\Pc      {\ensuremath{\mathrm{c}}\xspace}

 \def\Pi      {\ensuremath{\mathrm{i}}\xspace}

}
{

 \def\Ppi         {\ensuremath{\pi}\xspace}

 \def\Ppsi        {\ensuremath{\psi}\xspace}                 
                  
 \mathchardef\PDelta="7101
 \mathchardef\PXi="7104
 \mathchardef\PLambda="7103
 \mathchardef\PSigma="7106
 \mathchardef\POmega="710A
 \mathchardef\PUpsilon="7107
                  
 \def\PB      {\ensuremath{B}\xspace}                 
                  
 \def\PD      {\ensuremath{D}\xspace}

 \def\PJ      {\ensuremath{J}\xspace}                 
 \def\PK      {\ensuremath{K}\xspace}

 \def\PW      {\ensuremath{W}\xspace}

 \def\PZ      {\ensuremath{Z}\xspace}

 \def\Pc      {\ensuremath{c}\xspace}

 \def\Pi      {\ensuremath{i}\xspace}

}

\def\Wpm    {\ensuremath{\PW^\pm}\xspace}
\def\Z      {\ensuremath{\PZ^0}\xspace}

\def\c     {\ensuremath{\Pc}\xspace}

\def\pion  {\ensuremath{\Ppi}\xspace}

\def\pip   {\ensuremath{\pion^+}\xspace}
\def\pim   {\ensuremath{\pion^-}\xspace}

\def\kaon  {\ensuremath{\PK}\xspace}
  \def\Kbar  {\kern 0.2em\overline{\kern -0.2em \PK}{}\xspace}

\def\Kz    {\ensuremath{\kaon^0}\xspace}
\def\Kzb   {\ensuremath{\Kbar^0}\xspace}
\def\KzKzb {\ensuremath{\Kz \kern -0.16em \Kzb}\xspace}
\def\Kp    {\ensuremath{\kaon^+}\xspace}
\def\Km    {\ensuremath{\kaon^-}\xspace}

\def\KpKm  {\ensuremath{\Kp \kern -0.16em \Km}\xspace}

  \def\Dbar    {\kern 0.2em\overline{\kern -0.2em \PD}{}\xspace}
\def\D       {\ensuremath{\PD}\xspace}

\def\Dz      {\ensuremath{\D^0}\xspace}
\def\Dzb     {\ensuremath{\Dbar^0}\xspace}
\def\DzDzb   {\ensuremath{\Dz {\kern -0.16em \Dzb}}\xspace}
\def\Dp      {\ensuremath{\D^+}\xspace}
\def\Dm      {\ensuremath{\D^-}\xspace}

\def\DpDm    {\ensuremath{\Dp {\kern -0.16em \Dm}}\xspace}

\def\B       {\ensuremath{\PB}\xspace}
  \def\Bbar    {\kern 0.18em\overline{\kern -0.18em \PB}{}\xspace}

\def\Bu      {\ensuremath{\B^+}\xspace}
\def\Bub     {\ensuremath{\B^-}\xspace}

\def\Bd      {\ensuremath{\B^0}\xspace}

\def\jpsi     {\ensuremath{{\PJ\mskip -3mu/\mskip -2mu\Ppsi\mskip 2mu}}\xspace}
\def\psitwos  {\ensuremath{\Ppsi{(2S)}}\xspace}

  \def\Y#1S{\ensuremath{\PUpsilon{(#1S)}}\xspace}% no space before {...}!

\newcommand{\decay}[2]{\ensuremath{#1\!\to #2}\xspace}         % {\Pa}{\Pb \Pc}

\def\to                 {\ensuremath{\rightarrow}\xspace}

\def\BuToJPsiK  {\decay{\Bu}{\jpsi\Kp}}
\def\BdToDpi  {\decay{\Bd}{\Dm\pip}}
\def\BuToDpi  {\decay{\Bub}{\Dz\pim}}

\def\AT#1     {\ensuremath{A_T^{#1}}\xspace}           % 2

\def\C#1      {\ensuremath{\mathcal{C}_{#1}}\xspace}                       % 9
\def\Cp#1     {\ensuremath{\mathcal{C}_{#1}^{'}}\xspace}                    % 7
\def\Ceff#1   {\ensuremath{\mathcal{C}_{#1}^{\mathrm{(eff)}}}\xspace}        % 9  
\def\Cpeff#1  {\ensuremath{\mathcal{C}_{#1}^{'\mathrm{(eff)}}}\xspace}       % 7
\def\Ope#1    {\ensuremath{\mathcal{O}_{#1}}\xspace}                       % 2
\def\Opep#1   {\ensuremath{\mathcal{O}_{#1}^{'}}\xspace}                    % 7

\newcommand{\tev}{\ensuremath{\mathrm{\,Te\kern -0.1em V}}\xspace}
\newcommand{\gev}{\ensuremath{\mathrm{\,Ge\kern -0.1em V}}\xspace}
\newcommand{\mev}{\ensuremath{\mathrm{\,Me\kern -0.1em V}}\xspace}
\newcommand{\kev}{\ensuremath{\mathrm{\,ke\kern -0.1em V}}\xspace}
\newcommand{\ev}{\ensuremath{\mathrm{\,e\kern -0.1em V}}\xspace}
\newcommand{\gevc}{\ensuremath{{\mathrm{\,Ge\kern -0.1em V\!/}c}}\xspace}
\newcommand{\mevc}{\ensuremath{{\mathrm{\,Me\kern -0.1em V\!/}c}}\xspace}
\newcommand{\gevcc}{\ensuremath{{\mathrm{\,Ge\kern -0.1em V\!/}c^2}}\xspace}
\newcommand{\gevgevcccc}{\ensuremath{{\mathrm{\,Ge\kern -0.1em V^2\!/}c^4}}\xspace}
\newcommand{\mevcc}{\ensuremath{{\mathrm{\,Me\kern -0.1em V\!/}c^2}}\xspace}

\newcommand{\chisq}{\ensuremath{\chi^2}\xspace}

\def\gsim{{~\raise.15em\hbox{$>$}\kern-.85em
          \lower.35em\hbox{$\sim$}~}\xspace}
\def\lsim{{~\raise.15em\hbox{$<$}\kern-.85em
          \lower.35em\hbox{$\sim$}~}\xspace}

\def\ptot       {\mbox{$p$}\xspace}
\def\pt         {\mbox{$p_T$}\xspace}
\def\et         {\mbox{$E_T$}\xspace}

\def\tell1  {TELL1\xspace}
\def\ukl1   {UKL1\xspace}

\usepackage[T1]{fontenc} % all characters are made available but causes Type 3 font problems (in body text)
\usepackage{textcomp}
\begin{document}
\section{Introduction}
\begin{figure}[h]
\centering
\includegraphics[width=1.\columnwidth,clip=]{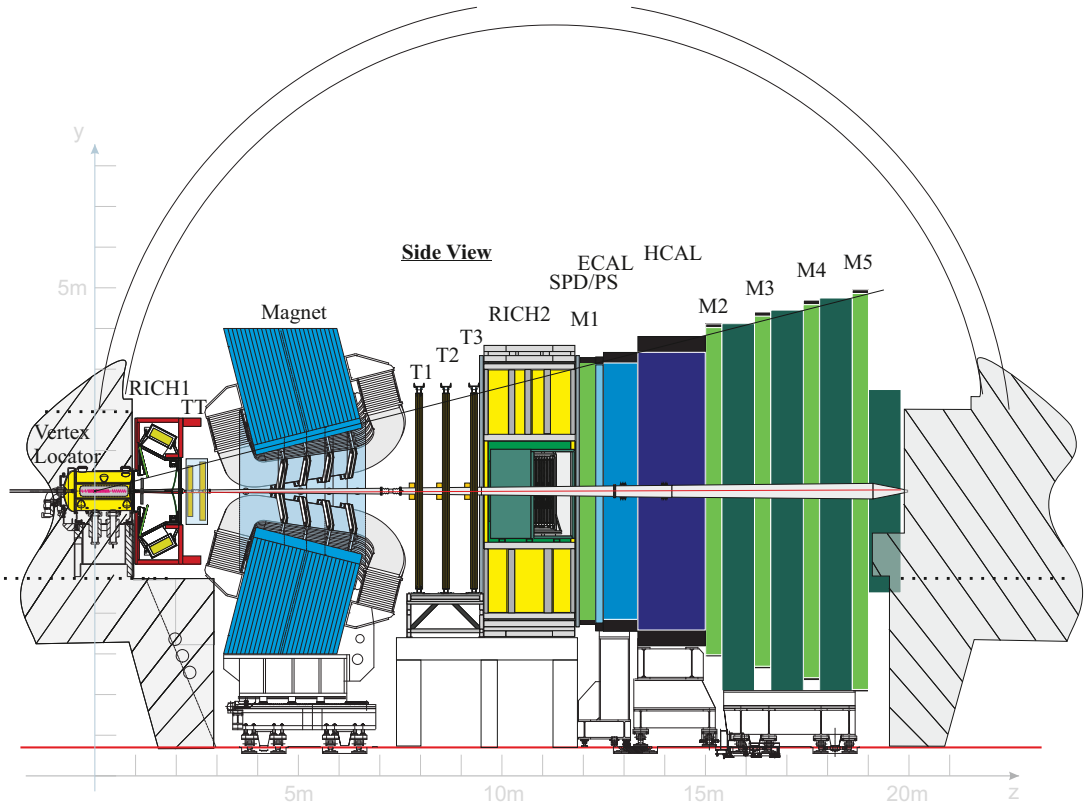}
\caption{Layout of the LHCb detector.} 
\label{layout}
\end{figure}

The LHCb detector~\cite{trigger} is a single-arm spectrometer that has been optimised to perform flavour
physics measurements at the LHC. LHCb has a pseudorapidity acceptance of
$2<\eta <5$. 
The detector layout is shown in Fig.~\ref{layout}. 
It consists of a silicon vertex detector surrounding the $pp$ interaction region (VELO); a silicon
strip detector (TT); a dipole magnet;
two Ring Imaging Cherenkov detectors (RICH 1$\&$2); 
tracking detectors (T1-T3), which consist of silicon strip detectors (IT) near the beam and straw
tubes (OT) further out;
a calorimeter system consisting of a Scintillating Pad detector (SPD), an electromagnetic calorimeter with 
pre-shower (ECAL, PS) and a hadronic calorimeter (HCAL); and muon chambers (M1-M5).
 
The LHCb trigger uses all of the above sub-systems. Its architecture consists of two levels, the first level trigger (L0) and
the High Level Trigger (HLT).
L0 is implemented in hardware and uses input from the calorimeter and muon systems.
L0 reduces the rate of crossings with at least one inelastic $pp$ interaction to below
1.1\,MHz, at which the whole detector can be read out. This maximum rate is imposed 
by the front-end (FE) electronics.
The implementation of L0 is only described briefly in Section~\ref{sec:l0}; a
fuller treatment can be found in reference~\cite{trigger}.
The HLT consists of a software application that runs on a farm of Personal Computers (PCs).
It has evolved significantly compared to reference~\cite{trigger}, 
in which it is assumed that the LHC machine would
operate with a 25\,ns bunch separation~\cite{lhc}, and that LHCb would limit the number of visible 
$pp$ interactions\footnote{A visible interaction is defined as one in which at least two tracks are
reconstructed in the VELO, that both point
to the interaction envelope.}
such that the average number of visible interactions per bunch crossing $\mu\simeq 0.4$.  
However, the smallest bunch separation of the machine was 50\,ns in the 2011 physics runs. 
To compensate for the loss in number of bunches, and combined with the fact that the LHCb detector performance did not degrade up to $\mu\approx 2.5$, LHCb decided to run 
at $\mu\approx 1.4$.  
Therefore the HLT had to adapt to running conditions rather different from those described in reference~\cite{trigger}.

The HLT is described in detail in Section~\ref{sec:hlt}. The HLT reduces the rate of accepted events to
$\sim 3$\,kHz, and all such events are written to storage.
The events written to storage are processed with a more accurate alignment and calibration of the sub-detectors,
and with reconstruction software that is more elaborate and allows for more redundancy than 
is possible in the HLT. This part of the reconstruction and subsequent event selection
will henceforth be referred to as the off-line reconstruction and selection.

The method used to obtain a data-driven determination of the trigger
performance is described in Section~\ref{sec:tistos}.
Section~\ref{sec:performance} describes the performance of the trigger in 2011 relative to off-line 
reconstruction and selection. 
Section~\ref{sec:conclusions} concludes with a summary of the trigger performance.
\section{First Level Trigger}
\label{sec:l0}
L0 is divided into three independent triggers; the L0-Calorimeter trigger, the L0-Muon trigger and the L0-PileUp trigger.
The last of these is not used to select flavour physics events, but instead aids the determination 
of the luminosity~\cite{Jaap}, and will not be further described in this paper. 

The L0 system is fully synchronous with the 40\,MHz bunch crossing signal of the 
LHC. The latencies are fixed and depend neither on the occupancy nor on the bunch crossing history. 
All of the L0 electronics are implemented in custom-designed hardware that makes use
of parallelism and pipelining to do the necessary calculations within the maximum latency of $4\,\mu$s.
The trigger decisions are combined in a single L0 decision, which
is transferred to the Readout Supervisor board (RS). The RS 
generates in addition a small rate of random (NoBias) triggers taking into account the bunch filling scheme of the machine\footnote{Not all of the 3564 slots available for proton bunches around the machine are
filled with protons. Most of the luminosity in 2011 was collected with 1296 bunches colliding
in LHCb.}.
The RS emulates the state of the FE buffers to protect against their overflow.
It also has information on the state of the buffers 
in the readout boards of all sub-detectors and the availability of the PCs in the farm. 
Based on this information it can retain or throttle a bunch crossing. 
\subsection{L0-Calorimeter Trigger Implementation}
\label{sec:l0caloimpl}
The L0-Calorimeter system uses information from the 
SPD, PS, ECAL and HCAL. These four detectors are stacked along the beam axis
($z$-axis) and their longitudinal segmentation offers the possibility to distinguish between photon,
 electron and hadron showers.
Transverse to the beam axis ($x$-$y$\,plane) the detectors are segmented into square cells. SPD, PS and ECAL are divided into three zones
with ECAL cell rib sizes of 40.4\,mm in the inner zone close to the beam pipe, 60.6\,mm and 121.2\,mm further out. HCAL is divided into two zones with rib size 131.3\,mm and 262.6\,mm. 
The SPD, PS and ECAL have the
same geometry and are projective, i.e. the sizes of the cells in the SPD and PS are adjusted to take into
account the different $z$ positions of the detectors. The HCAL cells are larger
but their boundaries always correspond to the boundaries of the ECAL cells.
The L0-Calorimeter system computes the transverse energy deposited in clusters of $2\times 2$ cells, 
using only cells located in the same zone. 
Hence the cluster energy of showers with energy deposits in two adjacent zones will be too low.
The transverse energy of a cluster is defined as:
\begin{equation}
      \et=\sum_{i=1}^4 E_i \rm{sin} \theta_{\it i} \,,
\label{trig-eq1}
\end{equation}
where $E_i$ is the energy deposited in cell $i$ and $\theta_i$ is the angle between the $z$-axis
and a neutral particle assumed to be coming from the mean position of the interaction envelope  
hitting the centre of the cell.
The ECAL and HCAL signals are read out and processed in FE boards (FEB) that cover
an area of $(8+1)\times (4+1)$ cells, such that the (+1) cells are shared between neighbouring FEB.  
Each FEB selects the highest \et cluster among its 32 clusters. From these clusters, three
 types of candidates are built combining information as follows:
\begin{enumerate}
\item Hadron candidate ({\tt L0Hadron}): the highest \et HCAL cluster. If there is a highest \et ECAL cluster located in front 
of the HCAL cluster, the \et of the hadron candidate is the sum of the \et of the
HCAL and ECAL clusters. 
\item Photon candidate ({\tt L0Photon}): the highest  \et ECAL cluster with 1 or 2 PS cells hit in front of the ECAL cluster 
and no hit in the SPD cells corresponding to the PS cells. In the inner zone of the ECAL,
an ECAL cluster with 3 or 4 PS cells hit in front of it is also accepted as photon.
The \et of the candidate is the \et deposited in the ECAL alone.
\item Electron candidate ({\tt L0Electron}): same requirements as for a photon candidate, with in addition at least one SPD cell hit in front of the PS cells. 
\end{enumerate}
The \et of the candidates is compared to a fixed threshold and events containing at least 
one candidate above threshold are retained by L0. 
\subsection{L0-Muon Trigger Implementation}
The muon system contains five muon stations (M1-M5) consisting of pads in the high occupancy regions and horizontal and vertical strips elsewhere.
Strips are combined to form logical pads for the muon trigger.
The pad sizes are chosen to obtain projectivity towards the interaction region in the $y$-$z$\,plane. 
Each quadrant of the muon detector is connected to a L0 muon processor. There is no exchange
of information between quadrants, hence muons traversing quadrant boundaries cannot be reconstructed in the trigger.
Each of the four L0 muon processors tries to identify the two muon tracks with the largest and second largest momentum transverse to the $z$-axis (\pt) in their quadrant.
The processors search for hits that define a straight line through the five muon stations and that
points towards the 
interaction point in the $y$-$z$\,plane. In the $x$-$z$\,plane the search is limited to muons with
$\pt\gtrsim 0.5$\,\gevc.
The position of a track in the first two stations allows the determination of its \pt with
a measured resolution of $\sim 25\,\%$ relative to off-line reconstructed muon tracks.
The trigger sets a single threshold on either the largest $\pt^{\rm largest}$ of the eight
candidates ({\tt L0Muon}), or a threshold on $\pt^{\rm largest}\times \pt^{\rm 2nd~largest}$ ({\tt L0DiMuon}).
\section{High Level Trigger}
\label{sec:hlt}
The HLT runs on the Event Filter Farm (EFF) that is a farm of multiprocessor PCs. The HLT is a program written in
C++, and 26110 copies of it run in the EFF. 
An event that is accepted by L0 is transferred by the on-line system from the FEB to the EFF
and is assembled by one of the event builder programs that run on one of the cores of each multicore node. 
The assembled events are placed in a buffer that is accessed by the HLT programs that run on the cores of the node.
A detailed description of this process can be found in reference~\cite{trigger} 
and references therein. 

The HLT is based on the same software as used throughout LHCb data processing and simulation~\cite{Corti}. The off-line event 
reconstruction and selection requires about 2\,s per event. During 2011 the L0 rate was about
870\,kHz. 
Given the available resources in the EFF this limits the time per event in the HLT to $\sim$30\,ms. 
The HLT is divided into two stages.
The first stage (HLT1) processes the full L0 rate and uses partial event reconstruction
to reduce the rate to 43\,kHz.
At this rate the second stage (HLT2) performs a more complete event reconstruction.

A "trigger line" is composed of a sequence of reconstruction algorithms and
selections. The trigger line returns an accept or reject decision. An event will be accepted by L0,
HLT1 or HLT2 if it is accepted by at least one of its trigger lines at the relevant stage. Combinations of trigger lines, together with a L0 configuration,
form a unique trigger with its associated Trigger Configuration Key (TCK). The TCK is a 32 bit label pointing to a database that contains the parameters that configure the trigger 
lines. The TCK is
stored for every event in the raw data, together with information on which trigger lines accepted the event. 
During 2011 running, the HLT contained 38 HLT1 and
131 HLT2 lines. The trigger lines that cover the main physics goals of LHCb~\cite{roadmap}, and
accept the majority of events stored,  are
described below in addition to the common reconstruction algorithms. The
corresponding selection parameters and their performance are given in Section~\ref{sec:performance}. 

The remaining trigger lines consist of lines for luminosity measurements, pre-scaled physics trigger lines
with looser cuts,  lines that select very low multiplicity events and lines that identify
large transverse momentum jets. 
The trigger also contains lines designed to accept NoBias events, lines that monitor events with inconsistent 
raw data or other errors during the HLT processing, lines that allow the VELO to monitor the 
position of the $pp$ interaction envelope and lines selecting calibration and monitoring data for fast feedback
on the quality of the data.
\subsection{HLT1}
\label{sec:HLT1}
The off-line VELO reconstruction software is fast enough to permit the full 3D pattern recognition of all
events that enter the HLT. In the off-line VELO pattern recognition a second pass is made on unused hits to
enhance the efficiency for tracks that point far away from the beam-line, but in the HLT this search is not
executed. 
At the start of each LHC fill, the mean position of the $pp$ interaction envelope in the $x$-$y$\,plane,
$\rm{PV}_{\rm{xy}}^{\rm{mean}}$, is 
determined using VELO tracks.
This position is measured to be stable to 
within a few $\rm\mu m$ per fill. 
The VELO tracks are used to construct vertices with at least 5 tracks originating from them, and
those vertices within a radius of $300~\mu\rm m$ of $\rm PV_{xy}^{mean}$ are considered to be primary vertices (PV).

While in the off-line pattern recognition all VELO tracks are considered to identify the corresponding
hits in the tracking stations downstream of the magnet, the pattern recognition in HLT1 limits the execution time by selecting VELO tracks 
that have a larger probability to originate from signal decays. 
HLT1 lines that do not require muons select VELO tracks based on their smallest impact parameter (IP) to any PV.  
In addition, cuts are applied to the quality of each VELO track based on the number of hits on a 
track and the
expected number of hits.

For events triggered by {\tt L0Muon} or {\tt L0DiMuon}, a fast muon identification is performed in 
HLT1 to select VELO tracks that are muon candidates using the following procedure. For every VELO
track, a search window is
defined in the M3 station by extrapolating the VELO track in a straight line.
The magnet does not bend tracks in the vertical plane, and multiple scattering 
dominates the vertical size of the search window.
A muon candidate is required to have a momentum of at least 6 \gevc, hence
the horizontal search window size corresponds to the deflection of a 6 \gevc track.
Hits found inside the search window are combined with the VELO track to form candidate
tracks that are used in a search for additional muon hits in stations M2, M4 and M5. 
A candidate track is provisionally accepted if
it contains at least one hit in addition to the M3 hit. In the final step of the algorithm,
a linear $\chi^2$ fit of the candidate track (containing both the VELO track and the muon hits) in
the horizontal plane is performed and the $\chi^2~\rm divided~by~the~number~of~degrees~of~freedom~(ndf)$ is required to be less than
25. As soon as the first candidate is found, the algorithm stops and the VELO track is tagged as a 
muon candidate.

For the VELO tracks that are selected by either their IP or by being tagged as a muon candidate, the
track-segments in the OT and IT-stations are reconstructed to determine their momentum in a procedure
known as forward tracking.
Imposing a minimum momentum and transverse momentum ($p,~\pt$) in the forward tracking significantly reduces the search windows that have to be opened in the IT and
OT tracking stations thereby reducing the required processing time.
Each reconstructed track is fitted using a Kalman filter \cite{Kalman} based track fit to obtain 
its $\chi^2$ 
and a full covariance matrix at the start of the track. Compared to off-line reconstruction,
this fit uses a simplified material geometry description, it makes fewer iterations and 
consequently it performs a less sophisticated removal of outlier hits.  
The invariant mass resolution of $\jpsi\rightarrow\mu^+\mu^-$ determined in the HLT is measured
to be $3\,\%$ larger than the 14\,\mevcc obtained off-line. This shows that the resolution of the track parameters obtained in
the HLT is sufficiently close to off-line to allow selective cuts in IP, momentum and mass.
For tracks that are tagged as muon candidates, the off-line muon identification algorithm \cite{ismuon}
is applied to the tracks to improve the purity of the muon sample.
\subsection{HLT2}
\label{sec:HLT2}
As mentioned above, HLT1 reduces the rate from 870 kHz to 43 kHz. At this rate forward tracking of all VELO tracks can be performed in HLT2.
While the off-line reconstruction uses two tracking algorithms, HLT2 only employs the algorithm based on seeding the search with VELO tracks. This leads to
a lower efficiency compared to off-line of $1-2~\%$ per track. 
To further limit the processing time only tracks with $p>5$\,\gevc and $\pt>0.5$\,\gevc are reconstructed 
by limiting the search windows.

Muon identification is performed using the off-line algorithm on all tracks from the forward
tracking. Tracks are also associated to ECAL clusters to identify electrons.

A large share of the 3\,kHz output rate of HLT2 is selected by "topological" trigger lines, 
which are designed to trigger on partially reconstructed $b$-hadron decays. 
The topological trigger lines in principle cover all $b$-hadrons with at least two charged particles in the final state and
a displaced decay vertex. The efficiencies are less
dependent on reconstruction inefficiencies imposed by the minimum ($p,~\pt$) requirements and loss
due to the single, non-redundant, track reconstruction mentioned above. 
In the following two sections
the topological trigger lines are described in more detail.

While the topological trigger lines target inclusive $b$-hadrons, a number of dedicated
"exclusive" trigger lines 
are also implemented in HLT2. These require all decay particles to be reconstructed in HLT2 and use narrow mass windows to reduce their rate.
These exclusive trigger lines either target prompt $c$-hadron production, or allow triggering on hadronic
$b$-hadron decays without the necessity to use lifetime-biasing selections to
reduce the rate. These lines are described in Section~\ref{sec:exclusive}.
\subsubsection{Topological Trigger Lines}
The decisions of the topological trigger lines are based on the properties of combinations of 2, 3, or 4 
``Topo-Tracks''.  Topo-Tracks are a subset of HLT2 tracks selected with additional requirements on their 
track fit quality ($\chi^2/\rm ndf$), IP, and muon or electron identification.
N-body (i.e. an n track combination) candidates are built as follows: two Topo-Tracks are combined into
a 2-body object, requiring that their distance of closest approach (DOCA) is less than 0.2\,mm.
A 3(4)-body object is made by combining a 2(3)-body object and another
Topo-Track with the same DOCA\,<0.2 mm cut, where the DOCA is calculated between 
the 2(3)-body object
and the additional Topo-Track.  This sequence of DOCA selections enhances the efficiency of the topological trigger lines on
$B\rightarrow DX$ decays.  
Not all of the $b$-hadron final state particles need to satisfy these criteria. The trigger is designed to allow
for the omission of one or more final state particles when forming the trigger candidate.

If an n-body candidate only contains a subset of the daughter particles, 
its invariant mass ($m$) will be less than the mass of a $b$-hadron. 
Thus, a mass window would need
to be very loose if the trigger is to be inclusive. 
Instead a corrected mass ($m_{\rm corr}$) is used that is defined as:

\begin{equation}
      m_{\rm corr}=\sqrt{m^2+|p^\prime_{T\rm miss}|^2}+|p^\prime_{T\rm miss}|,
\label{trig-eq2}
\end{equation}
where $p^\prime_{T\rm miss}$ is the missing momentum transverse to the direction of flight, 
as defined by the PV and the n-body vertex~\cite{mcorr}.
In case of multiple PVs, the PV with respect to which 
the n-body combination has the smallest IP is used.
The quantity $m_{\rm corr}$
would be the minimal mass of the parent if a massless particle was omitted from the trigger
candidate.
Prompt $c$-hadrons that are erroneously combined with another track constitute a significant fraction 
of the n-body candidates.
These candidates are rejected by requiring that
all (n-1)-body objects used by a n-body trigger line either have a mass greater than 2.5\,\gevcc
or that they have a significant IP to all PVs.

To select a n-body candidate, cuts are applied to the following variables:
$\sum|\pt|,~ p^{min}_{T}$, $m$, $m_{\rm corr}$, DOCA,  IP significance (IP$\chi^2$) and 
flight distance significance (FD$\chi^2$). 
Using NoBias events for background and Monte Carlo (MC) simulated signal events, we find that a larger
rejection power is achieved for the same signal efficiency by combining the above variables in a multivariate selection.
\subsubsection{Topological Multivariate Lines}
To combine the variables mentioned above a boosted decision tree (BDT) was chosen; this classifier has already been successfully
used elsewhere~\cite{BDT}.
All multivariate classifiers select 
n-dimensional regions of a multivariate space by learning from the training 
samples provided to them. 
If selected regions are small relative to the 
resolution of the detector, the signal could oscillate between
regions resulting in, at best, a less efficient trigger or, at worst, a trigger that
is very difficult to understand. 
To avoid this, all of the variables
are mapped onto discrete variables. The application of the BDT to discrete variables is henceforth referred to as
Bonsai BDT (BBDT).

The BBDT ensures that the smallest interval that can be used satisfies $\Delta x_{\rm min} > \delta_x$ 
for all $x$ values, where $\delta_x= {\rm MIN}\{ |x_i-x_j|:~x_i,~x_j \in x_{\rm discrete}\}$. 
The constraints governing the choice of $x_{\rm discrete}$ are then as follows:
firstly $\delta_x$ should be greater than the resolution on $x$ and be large with respect
to the expected variations in $x$, and secondly the discretisation should reflect common
$b$-hadron properties. 

The discretisation scheme for each variable was determined by first training a BBDT with a 
very large number of discretisation values and then gradually decreasing this number while maintaining near optimal performance. 
The training signal samples were MC simulated data 
that contained as signal $B^+, ~B^0, ~B_s$ or $\Lambda_b$ 
decays\footnote{Charge conjugate hadrons are always implied.} 
with decay modes as given in Table~\ref{tab:bbdtsamples}, 
while the background sample was NoBias data recorded in 2010. 
\begin{table}
\begin{center}  
  \caption{MC signal samples used to train the BBDT, where $K$ means $K^\pm$ and $\pi$ means $\pi^\pm$.}
  \label{tab:bbdtsamples} 
  \begin{tabular}{ c|c }
    Parent & Daughters \\ \hline
    $B^{+}$ & $K\pi\pi, D_{[K\pi]}\pi,D_{[hhhh]}K, D_{[K_S\pi\pi]}K,D_{[K\pi\pi]}K\pi$ \\
    $B^0$ & $K^*_{[K\pi]}\mu\mu,K^*_{[K\pi]}ee,D_{[K\pi\pi]}\pi,K\pi,D_{[K\pi]}K\pi,D^*_{[D(K\pi)\pi]}\mu\nu,D_{[K\pi\pi]}K\pi\pi$ \\
    $B_s$ &  $D_{s[KK\pi]} \pi, D_{s[KK\pi]}K\pi\pi,K^*_{[K\pi]}K^*_{[K\pi]}$ \\
    $\Lambda_b$ & $\Lambda_{c[pK\pi]} \pi,\Lambda_{c[pK\pi]} K\pi\pi$ \\
  \end{tabular}
  \end{center}
\end{table} 
Table~\ref{tab:bbdtdiscrete} shows the 
discretisation scheme for each of the variables used in the BBDT.
\begin{table}
\begin{center} 
    \caption{Allowed mapping points in the BBDT. The variables are explained in the text.}
    \label{tab:bbdtdiscrete} 
    \begin{tabular}{ c| c| c  }
    Variable & Cuts(2, 3, 4-body) & Intervals used in the BBDT \\ \hline
    $\sum |p_T|$ [\gevc] &  $>3$, 4, 4 & 3.5, 4, 4.5, 5, 6, 7, 8, 9, 10, 15, 20 \\
    $p_T^{\rm min}$ [\gevc]& $>0.5$ & 0.6, 0.7, 0.8, 0.9, 1, 1.25, 1.5, 1.75, 2, 2.5, 3, 4, 5, 10\\
    $m$ [\gevcc]& $< 7$ & 2.5, 4.75  \\
    $m_{\rm corr}$ [\gevcc] & & 2, 3, 4, 5, 6, 7, 8, 9, 10, 15 \\
    ${\rm DOCA}$ [mm] & $<0.2$& 0.05, 0.1, 0.15 \\
    ${\rm IP} \chi^2$ & & 20 \\
    ${\rm FD} \chi^2/100$ & $>1$ & 2, 3, 4, 5, 6, 7, 8, 9, 10, 25, 50, 100\\
  \end{tabular}
  \end{center}
\end{table} 
\subsubsection{Exclusive Lines}
\label{sec:exclusive}
In the topological trigger described in the previous section there is an explicit veto on prompt charm.
The selection of prompt charm decays is achieved by HLT2 lines that require a
reconstruction of all the decay products, and have tight cuts on the invariant mass of the
reconstructed candidates.
While HLT2 reconstruction and selection efficiencies for $b$-hadrons are good,
the $\pt>0.5$\,\gevc constraint reduces the efficiency for the exclusive selection of charm decays
with more than two final state particles. To enhance the reconstruction efficiency for these exclusive 
trigger lines, the trigger lines first try to identify a two-prong secondary vertex. Selection
cuts are imposed on the maximum invariant mass of the two tracks, the quality of their vertex, the sum of the
transverse momenta of the tracks and $m_{\rm corr}$. These initial cuts reduce the rate sufficiently to allow for 
the forward tracking of the remaining VELO tracks, but now with a relaxed $\pt>0.25$\,\gevc constraint and
only using the hits in the tracking system that have not been used by the previous pass in forward tracking.
Two-prong candidates are subsequently combined with other tracks, which now include the low \pt tracks, to form exclusively reconstructed
candidates. 
The combinatorial background is reduced by tight requirements on the mass and on the angle between the
momentum of the \D and the vector connecting the PV with the \D vertex.
HLT2 contains 28 trigger lines dedicated to selecting prompt charm.

Another example of an exclusive trigger is
a dedicated trigger line selecting the decay $B_s\rightarrow\,K^+K^-$ while avoiding cuts that bias the $B_s$ lifetime.
This implies that a cut on IP, a powerful variable to reject combinatorial background, cannot be used. In order to enrich the $B_s$ candidates two
dedicated neural networks based on the 
NeuroBayes neural network package~\cite{Feindt:2006pm} are used.
In a first step,  kinematic constraints such
as the transverse momentum of the final state particles
and the helicity angle in the rest frame of the $B_s$ candidate
are used to reduce the rate.
This allows running the comparatively slow 
particle identification algorithm using the RICH sub-detector
on the events selected by the first neural network.
This information is then included in a second neural
network that uses both kinematic and particle ID
information to make the final selection.
\section{Data-driven Trigger Performance Determination}
\label{sec:tistos}
The trigger performance is evaluated relative to offline reconstruction and selections, and
thus contains only the additional inefficiency due to simplifications used in the trigger, 
possible alignment inaccuracies, worse resolution than the offline reconstruction or harder cuts imposed by 
rate and/or processing time limitations.
The following channels have been chosen to show the performance of the trigger: 
$D^0\rightarrow K^-\pi^+$, $D^+\rightarrow K^-\pi^+\pi^+$,
$B^+\rightarrow\jpsi(\mu^+\mu^-)K^+$, 
$B^0\rightarrow D^-(K^+\pi^-\pi^-)\pi^+$,
$B^-\rightarrow D^0(K^-\pi^+)\pi^-$,
$B^0\rightarrow \jpsi(e^+e^-) K^{\ast 0}(K^+\pi^-)$ 
and $B^0\rightarrow K^{*0}(K^+\pi^-) \gamma$. 
These channels and their selections are representative for those
used in most analyses. In all off-line selected signal samples the level of background 
is significantly lower than the signal. 
Substantial differences in trigger efficiency, however, are observed for true signal and 
background.  
The trigger performance on each channel is measured by determining the signal component using fits to the invariant mass 
distributions, hence avoiding any background contamination. 

In what follows, the term ``signal'' refers to a combination of tracks that form
the off-line reconstructed and selected $b$ or $c$-hadron candidate. 
To determine the trigger efficiency, trigger objects are associated to signal tracks.
The trigger records all the information needed for such an association. 
All strips, straws, cells and pads of the sub-detectors have a unique identifier, and these
identifiers are written in a trigger report in the data stream for every trigger line that accepts an event.
The criteria used to associate a trigger object with
a signal track are as follows:
\begin{itemize}
\item L0-Calorimeter: the off-line track is extrapolated to the $z$-position of the calorimeter (ECAL or HCAL),
and the cells intersecting with the track and its eight neighbours are considered signal cells.
If any of the $2\times 2$ cells of a L0-Calorimeter 
cluster above the threshold coincides with a signal cell, this cluster is associated with the off-line track. If none of the cells overlap, the cluster is not associated with the off-line track.
\item L0-Muon: the trigger records the M1, M2 and M3 hits used to form the L0 muon candidate.
If at least two of the three hits are shared with an off-line reconstructed muon the L0 muon is
associated with the off-line track. 
Non-associated L0 muons have no hits overlapping between the L0 muon and the muon hits of the off-line track.
\item HLT tracks: a HLT track has VELO hits and hits in the OT and/or the IT. In addition it can have
TT hits and hits in the muon chambers M2-M5.  
Associated tracks require that the fraction of HLT track hits that overlaps with the off-line 
track is at least $70~\%$ in the VELO, $70~\%$ in the TT if applicable and $70~\%$ of the OT and IT combined. 
For muons the association requirement is that at least $60~\%$ of the HLT muon hits overlap
with the off-line muon. Non-associated HLT tracks share no hits with the off-line track.
\end{itemize}
An event is classified as TOS (Trigger on Signal) if the trigger objects that are associated with the signal are sufficient to trigger the event.
An event is classified as TIS (Trigger Independent of Signal) if it could have been triggered by those trigger objects that are not associated to the signal.
Global event variables, such as the number of primary vertices or the SPD multiplicity, are not
considered in this classification.
A number of events can be classified as TIS and TOS simultaneously ($N^{\rm TIS\& TOS}$), which allows
the extraction of the trigger efficiency relative to the off-line
reconstructed events from data alone.
The efficiency to trigger an event independently of the signal, $\epsilon\rm ^{\rm TIS}$,
is given by $\epsilon^{\rm TIS}=N^{\rm TIS\& TOS}/N^{\rm TOS}$, 
where $N^{\rm TOS}$ is the number of events classified as TOS. 
The efficiency to trigger an event on the signal alone, $\epsilon\rm ^{\rm TOS}$,
is given by $\epsilon^{\rm TOS}=N^{\rm TIS\& TOS}/N^{\rm TIS}$,
where $N^{\rm TIS}$ is the number of events classified as TIS. 
The total trigger efficiency for events containing the signal can then be
computed as $\epsilon^{\rm TIS} \times N^{\rm Trig}/N^{\rm TIS}$, where $N^{\rm Trig}$ is the total number of
triggered signal events.

\begin{figure}[h]
\centering
\includegraphics[width=0.5\columnwidth,clip=]{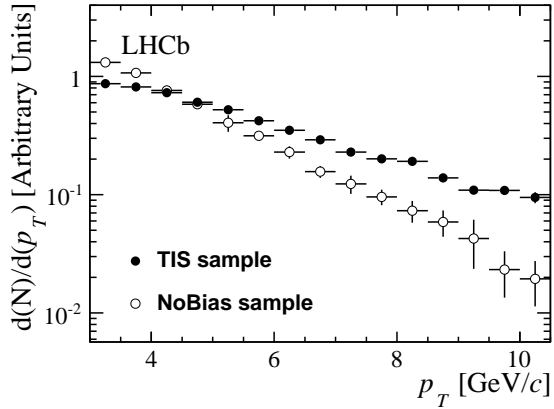}
\caption{Comparison of the \pt distributions of $D^+\rightarrow K^-\pi^+\pi^+$ selected in NoBias and TIS events.}
\label{fig:tistos}
\end{figure}

The phase-space distribution of the signal is affected by the TIS requirement. This is illustrated in Fig.~\ref{fig:tistos}, which
shows the \pt distribution of $D^+\rightarrow K^-\pi^+\pi^+$ candidates selected from NoBias and TIS events.
The \pt of TIS events is harder, which would result in a too large signal efficiency when integrated over all phase-space.
This bias in phase-space can be understood considering that $c\bar{c}$, or $b\bar{b}$, pairs are correlated in
phase-space at production, and TIS events are predominantly triggered by the decay 
products of the hadron that contains the other heavy quark. Another example of bias is that for charm decays  
TIS events could enhance the non prompt charm component by triggering on the other decay products of a \B to \D decay. These biases can only be evaluated individually for each analysis.
Therefore the trigger
performance is presented as a function of the signal $\pt$ and its lifetime ($\tau$). 

\section{Trigger Performance}
\label{sec:performance}
For each channel $\varepsilon^{\rm TOS}$ is determined relative to the off-line selection efficiency
of a channel. $\varepsilon^{\rm TOS}$ for the HLT1(2) performance is given for off-line selected events that have also been
classified as TOS in the previous trigger level(s), unless mentioned otherwise. 

At each trigger level
the different trigger lines compete for their share of the available resources. To determine
the different selections for the trigger lines, a "bandwidth division" procedure has been adopted, which is described in the next section. The performance of the different trigger lines with the thresholds as determined by the
bandwidth division will be presented for L0, HLT1 and HLT2 in the next sections for the channels listed in
Section~\ref{sec:tistos}. 
\subsection{Bandwidth Division Procedure}
The bandwidth division minimises the overall loss in efficiency
by minimising the following:
\begin{equation}
      \sum_{\rm signal} \left( \sum_{\rm lines}
         { \left( 1-\frac{\varepsilon^{\rm signal,~line}}
               {\varepsilon_{\rm max}^{\rm signal,~line}} \right) } \right)^2 ,
\end{equation}
where $\varepsilon^{\rm signal,~line}$ is the L0$\times$HLT trigger efficiency
obtained using a set of selections (corresponding to a single set of cut values) 
for all signal channels and trigger lines simultaneously and $\varepsilon_{\rm max }^{\rm signal,~line}$  
is the maximum of $\varepsilon^{\rm signal,~line}$ with the full computing resources dedicated to that signal 
and specific trigger line alone.
The score is evaluated for each set of cuts by running an emulation of the L0 trigger and 
executing
the HLT application. This emulation includes FE-buffer overflow emulation, the 
available processing power in the EFF and the maximal HLT2 output rate to disk as 
boundary conditions. The configuration with the minimum score is found by varying the cuts and running the
trigger software for each variant.

For signal the following MC generated and off-line reconstructed and selected channels have been chosen to represent both the 
main physics goals of LHCb and to cover all the trigger components that need to be tuned: 
$B_s\rightarrow J/\psi(\mu^+\mu^-)\phi(K^+K^-)$,
$B_s\rightarrow\mu^+\mu^-$,
$B^0\rightarrow K^{*0} \mu^+\mu^-$,
$B_{(s,d)}\rightarrow\mu^+X$,
$D^{*+}\rightarrow D(\mu^+\mu^-)\pi^+$,
$D^+\rightarrow K^-\pi^+\pi^+$,
$B^0\rightarrow K^{*0}\gamma$,
$B^0\rightarrow K^+\pi^-$,
$B^0\rightarrow K^{*0}e^-e^+$,
$B^+\rightarrow K^+\pi^-\pi^+$,
$B_s\rightarrow D_s^-(K^+K^-\pi^-)K^+$,
$B^+\rightarrow \bar{D^0}(K_S(\pi^+\pi^-)\pi^+\pi^-)K^+$ and
$D^0\rightarrow K^-\pi^+$ with $K^{\ast 0}\rightarrow K^+\pi^-$. 
NoBias events from 2010 with $\mu=1.4$ are used as background.
Rather than introducing weights favouring some channels, we have chosen to emphasize the main physics goals by the number of channels included in the bandwidth division procedure. 
For example, channels decaying with muons in the final
state are more abundant.

All performance results are given for 1296 colliding bunches in LHCb, which corresponds to a bunch crossing rate with
at least one visible $pp$ interaction of $\sim 11$\,MHz. 
The bandwidth division yields the following rates for NoBias events: 
870\,kHz for L0, 43\,kHz for HLT1 and 3\,kHz for HLT2.  
\subsection{L0 Performance}
Events with a large occupancy in the OT and IT consume a disproportionately 
large fraction of the available processing time in the HLT.
The SPD multiplicity measured at L0 is a good measure of this occupancy, 
permitting an early rejection of events
that require a relatively large processing time.
Using the bandwidth division the optimal SPD cut is determined to be $<~900$ for events
triggered by {\tt L0DiMuon} and  $<~600$ for 
all other L0 triggers. On average, events with a SPD multiplicity larger than 
600 consume four times more time in the HLT  
than events with less than 600 SPD hits.
The fraction of events rejected
due to these cuts has been determined from real data for charm hadron production to be
$7.4\pm~0.3~(0.05\pm 0.01)\,\%$ for a cut on 600 (900) in SPD multiplicity. 
Similarly for $b$-hadron production the fraction of events with a SPD multiplicity 
$>600~(900)$ is found to be
$8.8\pm0.6~(0.5\pm 0.2)\,\%$.
All efficiencies quoted below are given relative to the sample after the SPD 
multiplicity cut.

\begin{figure}
\begin{minipage}[t]{7.77cm}
\includegraphics[width=0.9\textwidth]{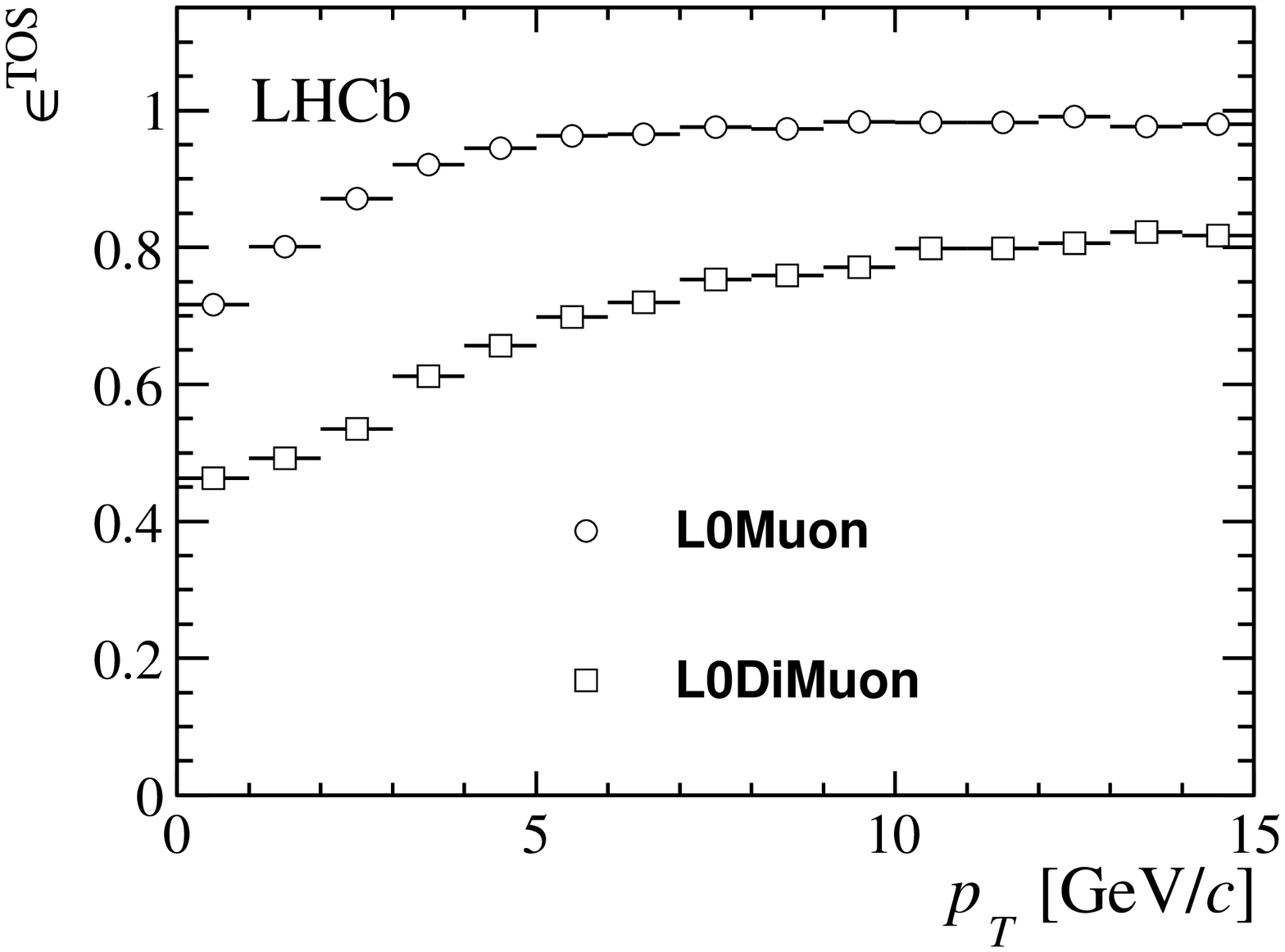}
  \caption{Efficiency $\epsilon^{\rm TOS}$ of $B^+\rightarrow \jpsi(\mu^+\mu^-)K^+$ as a function of \pt(\jpsi) for {\tt L0Muon} and {\tt L0DiMuon} lines.
}
  \label{f:l0muon}
\end{minipage} 
\hfill
\begin{minipage}[t]{6.23cm}
\includegraphics[width=0.9\textwidth]{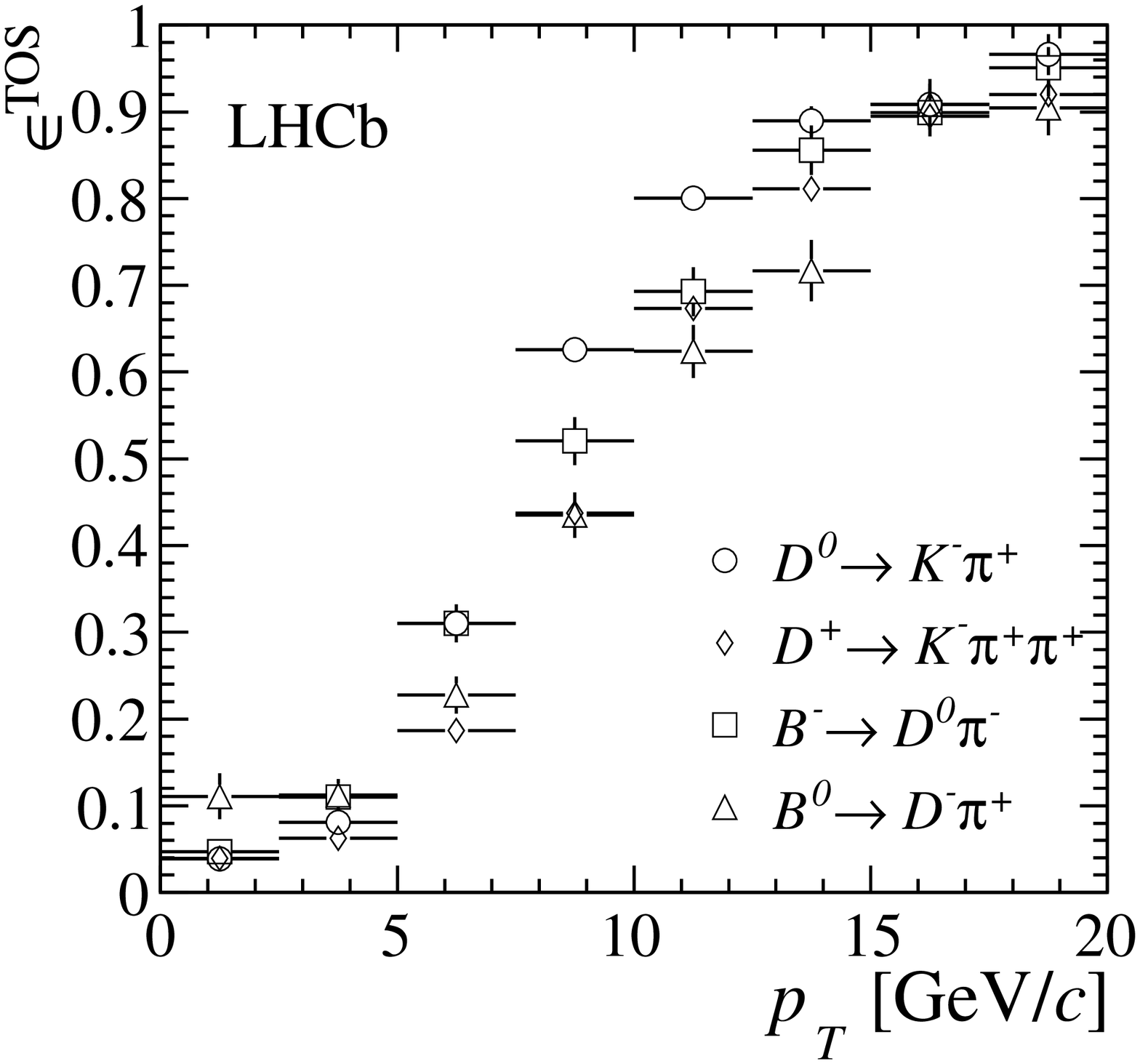}
\caption{The efficiency $\epsilon^{\rm TOS}$ of {\tt L0Hadron} is shown for \BdToDpi, 
\BuToDpi, 
$D^0\rightarrow K^-\pi^+$ and $D^+\rightarrow K^-\pi^+\pi^+$ as a 
function of \pt of the signal $B$ and $D$ mesons.
} 
\label{f:l0hadron}
\end{minipage}
\hfill
\end{figure}

Table~\ref{tab:l0-cuts} lists the L0 cuts. About $20~\%$ of the
events accepted by L0 are selected by more than one trigger line, 
giving a total L0 rate of 870 kHz prior to throttling. 
\begin{table*}[ht]
\begin{center}
\caption{Cuts of L0 lines and their rates prior to throttling. 
The definition of the trigger lines is given in Section~2.}
\label{tab:l0-cuts}
\begin{tabular}{l|r|r|r}
 &Threshold &SPD Multiplicity & Rate\\\hline
{\tt L0Muon}& $\pt>1.48$ \gevc & $<600$ & 340 kHz\\
{\tt L0DiMuon}& $\sqrt{\pt^{\rm largest}\times \pt^{\rm 2nd~largest}}>1.296$ \gevc & $<900$ & 75 kHz\\
{\tt L0Hadron}& $E_{T}>3.5$ \gev & $<600$ & 405 kHz\\
{\tt L0Electron}& $E_{T}>2.5$ \gev & $<600$ & 160 kHz\\
{\tt L0Photon}& $E_{T}>2.5$ \gev & $<600$ & 80 kHz\\
\end{tabular}
\end{center}
\end{table*}
{\tt L0Muon} is the main trigger for particle decays with one or more muons in the 
final state. {\tt L0DiMuon} 
recovers part of the events with a SPD multiplicity $>600$ for a small increase in rate.
The performance of {\tt L0Muon} and {\tt L0DiMuon} are shown in Fig.~\ref{f:l0muon} for
$B^+\rightarrow \jpsi(\mu^+\mu^-)K^+$ as a function of
\pt(\jpsi). 
{\tt L0DiMuon} increases the number of
signal events by $4.9~\%$, of which $87~\%$ have a SPD multiplicity larger than 600 hits. 
The remaining $13~\%$ is due to the lower \pt cut in {\tt L0DiMuon}.  
L0 requires a muon candidate to have a hit in all five muon stations, while off-line as few as two
stations are sufficient to identify a muon. 
As a result {\tt L0DiMuon} has a maximum efficiency of $\sim 80~\%$ even for a \jpsi with 
large \pt. {\tt L0Muon} recovers this
loss for lower SPD multiplicities and decays with more muons at large \pt.

{\tt L0Hadron} selects heavy flavour decays with hadrons in the final state.  
The performance of {\tt L0Hadron} is shown in Fig.~\ref{f:l0hadron} for \BdToDpi, \BuToDpi, 
$D^0\rightarrow K^-\pi^+$ and $D^+\rightarrow K^-\pi^+\pi^+$ as a 
function of \pt  of the signal $B$ and $D$ mesons.
At low \pt, {~\tt L0Hadron} has a better efficiency for $b$-hadrons than for $c$-hadrons due to the larger
$b$-hadron mass. Once the \pt of the hadron is above the $b$-hadron mass, the decays with fewer final state tracks have a higher efficiency. 

\begin{figure}[h]
\centering
\includegraphics[width=0.5\columnwidth,clip=]{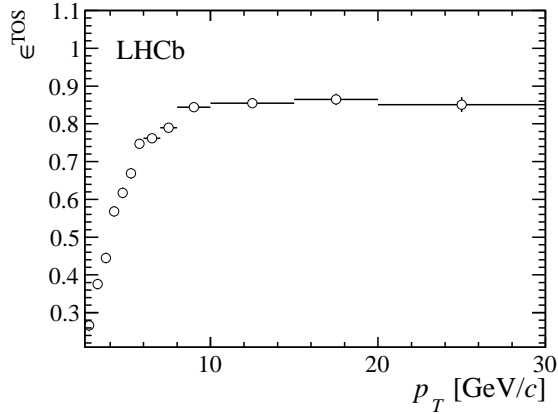}
\caption{The efficiency $\epsilon^{\rm TOS}$ of {\tt L0Electron} is shown for
${B}^0\rightarrow J/\psi{(\rm e^+e^-)}K^{*0}$ as a function of \pt(\jpsi).
} 
\label{f:l0electron}
\end{figure}

{\tt L0Electron} selects decays with electrons in the final state. It also triggers on radiative decays, with the photon being
either converted, or with photon clusters with SPD hits in front due to overlapping charged particles.
The performance of {\tt L0Electron} is shown in Fig.~\ref{f:l0electron} for ${B}^0\rightarrow J/\psi{(\rm e^+e^-)}K^{*0}$ as a 
function of $\pt(\jpsi)$. Contrary to {\tt L0Muon}, {\tt L0Electron} is not fully efficient for \jpsi with large \pt. This is due to the hardware implementation (see Section~\ref{sec:l0caloimpl}) that prevents
energy deposited in different ECAL zones from being combined into one cluster.

The performance of the trigger for high-energy photons from radiative penguin decays is measured with the
channel $B^0\rightarrow K^{*0} \gamma$. The
number of TIS events in this channel is insufficient to study the efficiency as a function of 
\pt of the $B^0$. The mean efficiency for the {\tt L0Photon} line integrated over \pt is $50\pm 4~\%$. 
Selecting events with either {\tt L0Photon} or {\tt L0Electron} gives an efficiency of
$88\pm 5~\%$ .
\subsection{HLT1 Performance}
HLT1 muon lines are only executed for events that have been triggered by {\tt L0Muon} or {\tt L0DiMuon}, and
the lines require their tracks to be validated as a muon candidate as described in Section~\ref{sec:HLT1}.
Table~\ref{t:hlt1muon} gives the names of the HLT1 muon lines and their cuts.
\begin{table*}[ht]
\caption{HLT1 muon lines and their cuts. The rate is measured on events accepted by {\tt L0Muon} or {\tt L0DiMuon}.
}
\label{t:hlt1muon}
\begin{center}
\begin{tabular}{l|r |r |r| r }
Hlt1line &{\tt TrackMuon}&{\tt SingleMuon}&{\tt DiMuon}&{\tt DiMuon}\\
 & &{\tt HighPT}&{\tt HighMass}&{\tt LowMass}\\\hline
Track IP [mm]&$>0.1$&-&-&-\\
Track IP$\chi^2$&$>16$&-&-&$>3$\\
Track \pt [\gevc]& $>1$&$>4.8$&$>0.5$&$>0.5$\\
Track \ptot [\gevc]& $>8$&$>8$&$>6$&$>6$\\
Track $\chi^2/\rm ndf$&$<2$&$<4$&$<4$&$<4$\\
DOCA [mm]&-&-&$<0.2$&$<0.2$\\
$\chi_{\rm vertex}^2$&-&-&$<25$&$<25$\\
Mass [\gevcc]&-&-&$>2.7$&$>1$\\\hline
Rate [kHz]&5&0.7&1.2&1.3\\
\end{tabular}
\end{center}
\end{table*}
\sloppy
{\tt Hlt1TrackMuon} accepts events with \B, \D or $\tau$ decays with at least one muon in its final state 
by identifying and accepting events with muon candidates that have significant IP with respect to all PVs. 
To trigger on muons originating from heavy 
particles with a negligible lifetime, like \Wpm or \Z, an alternative line, 
{\tt Hlt1SingleMuonHighPT}, is implemented. 
It does not have any requirements on IP, but requires a hard \pt
cut to reduce the rate. {\tt Hlt1DiMuonHighMass} is complementary to {\tt Hlt1TrackMuon} in that it allows 
$b$-hadron decays to be selected without imposing lifetime related cuts, 
and thus allows the
lifetime acceptance bias of the larger efficiency {\tt Hlt1TrackMuon} line to be determined.
Finally {\tt Hlt1DiMuonLowMass} allows triggering on final states with two muons with a relatively small invariant mass.
To reduce the rate the line requires that both muons are not prompt.

\begin{figure}[htbp]
\centering
\mbox{\subfigure{\epsfig{figure=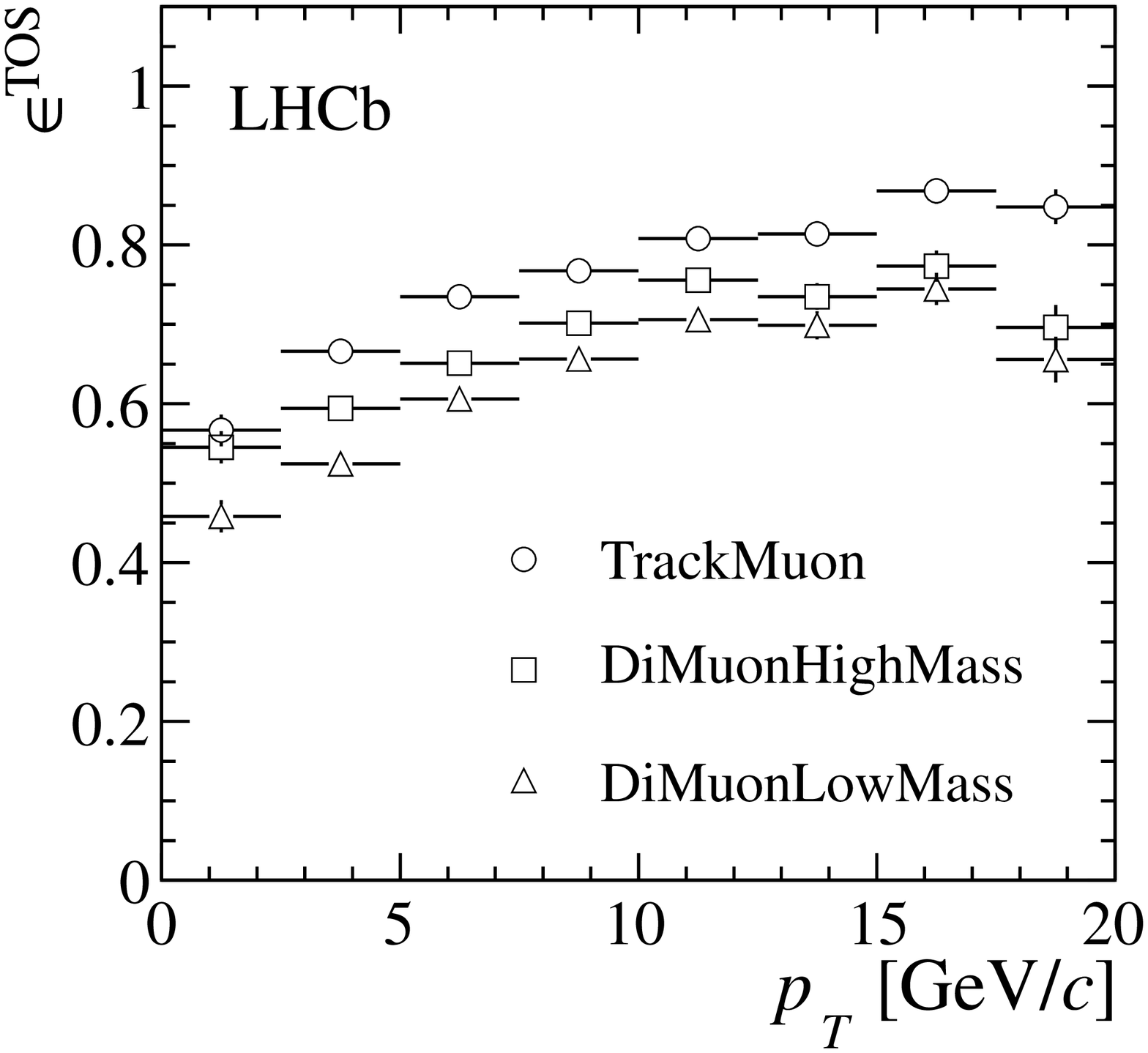,width=0.5\textwidth}}
\subfigure{\epsfig{figure=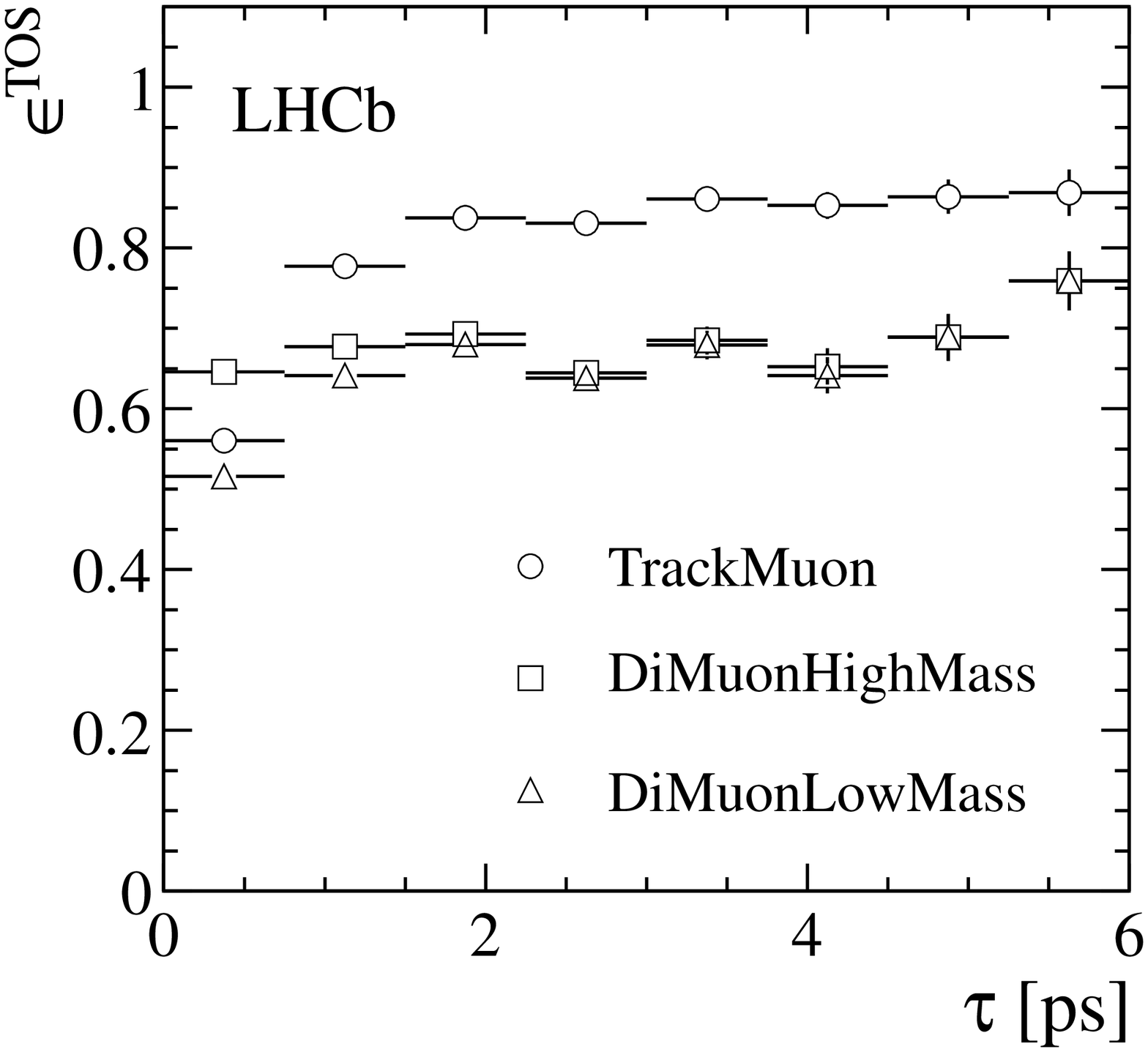,width=0.5\textwidth}}}
\caption{Efficiency $\epsilon^{\rm TOS}$ of {\tt Hlt1TrackMuon}, {\tt Hlt1DiMuonHighMass} 
and {\tt Hlt1DiMuonLowMass} for $B^+ \rightarrow \jpsi (\mu^+\mu^-) K^+$
as a function of the \pt and lifetime of the $B^+$. 
}
  \label{fig:hlt1muon}
\end{figure}

The performance of the HLT1 muon lines is evaluated using \BuToJPsiK decays.
Figure~\ref{fig:hlt1muon} shows the performance of {\tt Hlt1TrackMuon}, {\tt Hlt1DiMuonHighMass} and {\tt Hlt1DiMuonLowMass}
as a function of the \pt and $\tau$ of the $B^+$. 
{\tt Hlt1TrackMuon} gives the best performance overall, except at low lifetimes, 
where {\tt Hlt1DiMuonHighMass} recovers events. {\tt Hlt1DiMuonLowMass} loses $\sim 10~\%$ in efficiency
compared to {\tt Hlt1TrackMuon} for \BuToJPsiK due to the requirement to have at least two muon 
candidates, but its cuts on IP and \pt are significantly relaxed to allow the selection of candidates with
the muon pair mass down to 1\,\gevcc, which is designed to select  $b\rightarrow s\mu\mu$ decays like $B\rightarrow
K^{*}\mu^+\mu^-$.

\fussy
The performance of {\tt Hlt1SingleMuonHighPT} is not
properly assessed using \BuToJPsiK decays because it is designed to accept events with decays of particles with
a mass larger than that of $b$-hadrons.
Instead $Z^0~\rightarrow~\mu^+~\mu^-$ events are used to measure the efficiency, by requiring one of the two muons to 
be TIS. This yields an efficiency of $77.1\pm0.2~\%$
for the {\tt Hlt1SingleMuonHighPT} line per single muon, implying an efficiency for $Z^0~\rightarrow~\mu^+~\mu^-$ of $95~\%$. 
The fast HLT1 muon reconstruction, 
as described in Section~\ref{sec:HLT1}, 
applies more stringent cuts than the off-line muon identification to keep misidentification and as a consequence the rate and CPU time consumption under control.

\begin{figure}[h]
\centering
\mbox{\subfigure{\epsfig{figure=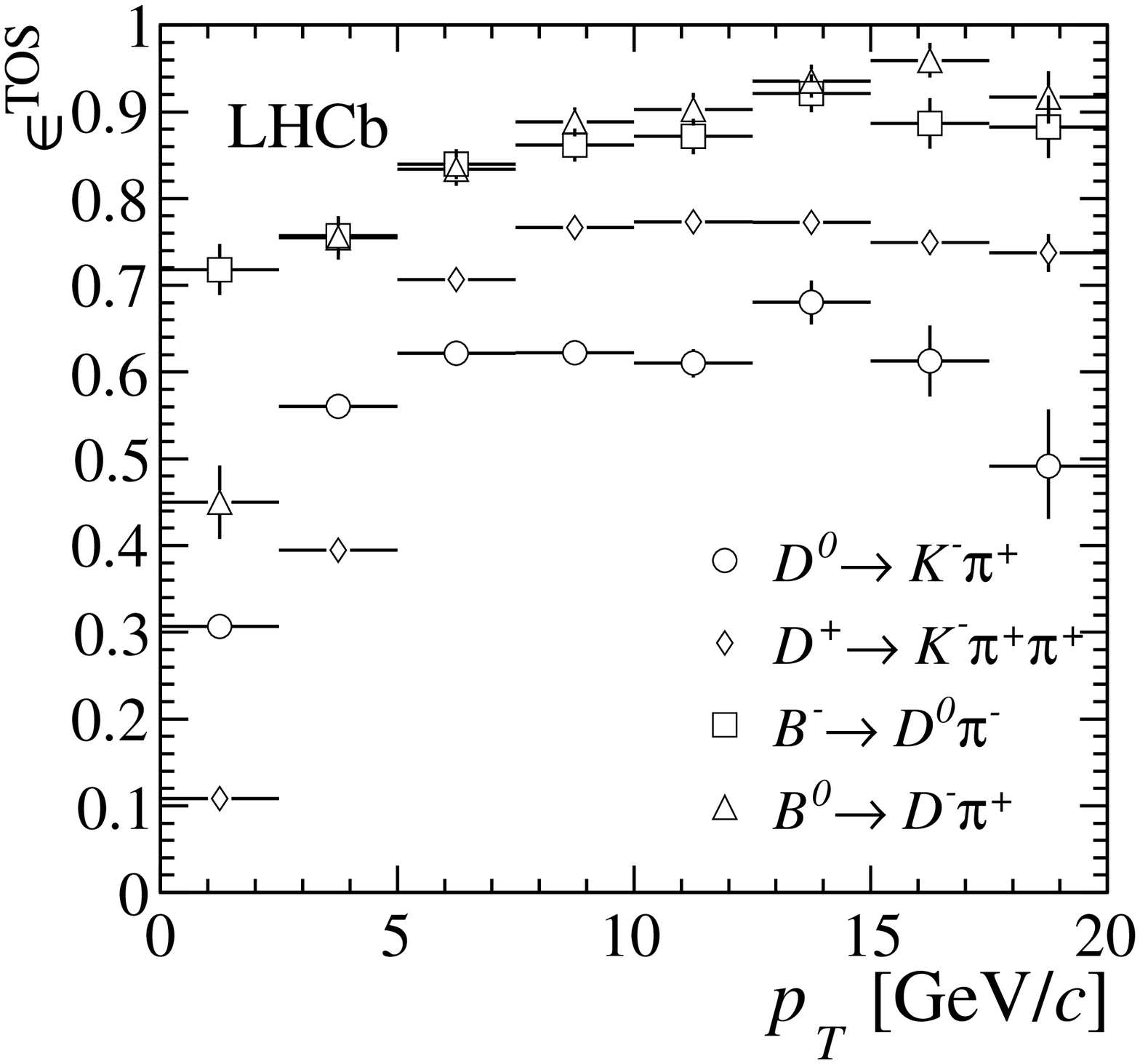,width=0.5\textwidth}}
\subfigure{\epsfig{figure=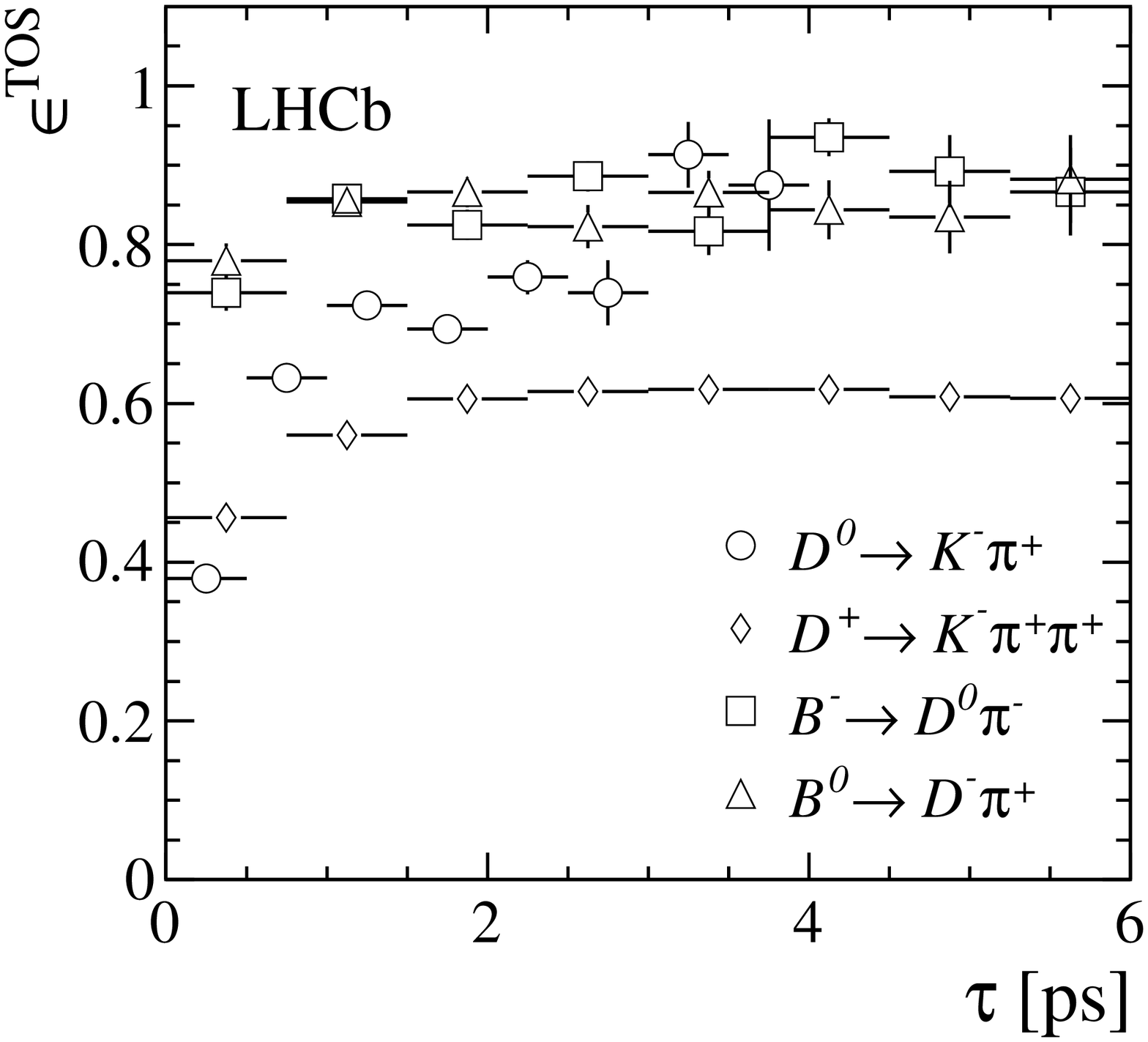,width=0.5\textwidth}}}
\caption{Efficiency $\epsilon^{\rm TOS}$  of {\tt Hlt1TrackAllL0} is shown
for \BuToDpi, \BdToDpi, $D^0\rightarrow K^-\pi^+$ and $D^+\rightarrow K^-\pi^+\pi^+$ as a 
function of \pt and  $\tau$ of the \B-meson and prompt \D-meson respectively.
} 
\label{fig:hlt1all}
\end{figure}

In addition to the muon lines mentioned above, HLT1 also contains a line that is executed for all events accepted by L0, {\tt Hlt1TrackAllL0}.
It is designed to select hadron decays which are significantly displaced from a PV.
A trigger line called {\tt Hlt1TrackPhoton} is only executed for events that have a {\tt
L0Photon} or a {\tt L0Electron} with $E_{T}>4.2$\,\gev. This trigger line is designed to enhance the trigger efficiency for radiative $b$-hadron decays with a high \pt photon.
The corresponding selection cuts are given in Table~\ref{tab:hlt1all}. Both trigger lines require at least one
track with sufficient IP and \pt. {\tt Hlt1TrackPhoton} is designed to select lower \pt tracks, and 
correspondingly also has relaxed track quality requirements compared to {\tt Hlt1TrackAllL0}. 
\begin{table*}[ht]
\caption{The cuts applied in {\tt Hlt1TrackAllL0} and {\tt Hlt1TrackPhoton} lines. The rate is measured on events accepted by
L0.}
\label{tab:hlt1all}
\begin{center}
\begin{tabular}{l|r |r }
Hlt1 line &{\tt Hlt1TrackAllL0}&{\tt Hlt1TrackPhoton}\\\hline
Track IP [mm]&$>0.1$&$>$0.1\\
Number VELO hits/track&$>9$&$>6$\\
Number missed VELO hits/track&$<3$&$<3$\\
Number OT+IT$\times$2 hits/track&$>16$&$>15$\\
Track IP$\chi^2$&$>16$&$>$16\\
Track \pt [\gevc]& $>1.7$&$>1.2$\\
Track \ptot [\gevc]& $>10$&$>6$\\
Track $\chi^2/\rm ndf$&$<2.5$&$<2.5$\\
\hline
Rate [kHz]&33&4.2\\
\end{tabular}
\end{center}
\end{table*}
Figure~\ref{fig:hlt1all} shows the performance of {\tt Hlt1TrackAllL0} as a function of $\pt\rm~and~\tau$ for channels with hadronic decays. 
{\tt Hlt1TrackAllL0} provides a very efficient trigger for all heavy flavour decays with a significant flight distance from
their PV, reducing the rate from 870\,kHz to 33\,kHz. At low \pt the requirement of at
least one decay particle with \pt$>1.7$\,\gevc results in 
selecting $b$-hadrons with a larger efficiency than $c$-hadrons, and low multiplicity decays 
with a larger efficiency than higher multiplicity decays. 
At large \pt this condition favours the decays with larger multiplicities.
While in L0 special lines are used to select $b$-hadron decays with electrons
in the final state, in HLT1 these decays are covered by {\tt Hlt1TrackAllL0}.

\sloppy
There are insufficient radiative \B-decays to extract the performance of 
{\tt Hlt1TrackPhoton} as a function of \pt
in a data-driven way. {\tt Hlt1TrackPhoton} uses the same tracks as {\tt Hlt1TrackAllL0}, 
but with a relaxed set of requirements 
as shown in Table~\ref{tab:hlt1all}. 
The yield increase in $B^0\rightarrow K^{*0}\gamma$ events obtained by 
including {\tt Hlt1TrackPhoton} in addition to {\tt Hlt1TrackAllL0} is measured to be $12\pm 2~\%$.

\fussy
\subsection{HLT2 Performance}
Similar to HLT1, HLT2 has lines that select events with one or two identified muons in the final state. 
In HLT2 the
muon identification is identical to the off-line algorithm.
The cuts corresponding to lines that are purely based on a single identified muon are given in Table~\ref{tab:hlt2singlemuon}.
\begin{table*}[ht]
\caption{HLT2 lines based on one identified muon.
}
\label{tab:hlt2singlemuon}
\begin{center}
\begin{tabular}{l|r | r }
{\tt Hlt2Single} &{\tt Muon}&{\tt MuonHighPT}\\\hline
{\tt Hlt1TrackMuon}&TOS&-\\
Track IP [mm]&$>0.5$&-\\
Track IP$\chi^2$&$>200$&-\\
Track \pt [\gevc]& $>1.3$&$>$10\\
Track $\chi^2/\rm ndf$&$<2$&-\\\hline
Pre-scale&0.5&1.\\
Rate [Hz]&480&45\\
\end{tabular}
\end{center}
\end{table*}
{\tt Hlt2SingleMuon} selects semileptonic $b$ and $c$-hadron decays.
To minimise the bias on the hadronic part of the decay the \pt cut is set low, in combination with scaling the rate down by a factor two, rather than tightening the cut to reduce the rate.
This trigger line also provides a large yield for $J/\psi\rightarrow\mu\mu$ events that are selected by
one of the two muons, while the other muon is used for calibration of tracking and muon
identification efficiencies. 
{\tt Hlt2SingleMuonHighPT} is designed to select heavy particles 
decaying promptly to one or more muons, like \Wpm or \Z. Contrary to {\tt Hlt2SingleMuon} the rate is
not a problem so there is no HLT1 requirement imposed. As for HLT1, $Z^{0}\rightarrow\mu^{+}\mu^{-}$ decays are used to measure an efficiency of over $99\%$ for {\tt Hlt2SingleMuonHighPT} per muon.
This small loss in efficiency is attributed to different alignment constants and the non-redundant track reconstruction used in HLT2 as described in Section~\ref{sec:HLT2}. 

\sloppy
The HLT2 lines that are based on two identified muons are grouped into two categories. Those
that are dedicated to prompt decays use the mass as the main discriminant, 
while "detached" lines use the separation between the dimuon vertex and the PV as the 
main discriminant.
The names and corresponding cuts of the prompt decay selections are given in 
Table~\ref{tab:hlt2dimu}.
\begin{table*}[ht]
\caption{HLT2 lines based on two identified muon.
}
\label{tab:hlt2dimu}
\begin{center}
\begin{tabular}{l|r |r |r| r|r }
{\tt Hlt2DiMuon} &{\tt JPsi}&{\tt Psi2S}&{\tt B}&{\tt JPsiHighPT}&{\tt Psi2SHighPT}\\\hline
Track $\chi^2/\rm ndf$&$<5$&$<5$&$<5$&$<5$&$<5$\\
Mass [\gevcc]&M$_{J/\psi}\pm 0.12$&M$_{\psitwos}\pm 0.12$&$>4.7$&M$_{J/\psi}\pm 0.12$&M$_{\psitwos}\pm 0.12$\\
$\chi_{\rm vertex}^2$&$<25$&$<25$&$<10$&$<25$&$<25$\\
$\pt^{\mu\mu}$ [\gevc]&-&-&-&$>2$&$>3.5$\\\hline
Pre-scale&0.2&0.1&1.&1.&1.\\
Rate [Hz]&50&5&80&115&15\\
\end{tabular}
\end{center}
\end{table*}
{\tt Hlt2DiMuonJPsi(Psi2S)} and {\tt Hlt2DiMuonJPsi(Psi2S)HighPT} all select a mass region 
around $\jpsi~(\psitwos)$. 
{\tt Hlt2DiMuonJPsi(Psi2S)} avoids explicit \pt requirements but as a consequence needs to 
be 
pre-scaled to reduce the rate. {\tt Hlt2DiMuonJPsi(Psi2S)HighPT} reduces the prompt 
$\jpsi~(\psitwos)$ rate
by applying a \pt cut on the $\jpsi~(\psitwos)$ candidate. {\tt Hlt2DiMuonB} has 
its mass cut set
high enough to have an acceptable rate.

\begin{figure}[htbp]
\centering
\mbox{\subfigure{\epsfig{figure=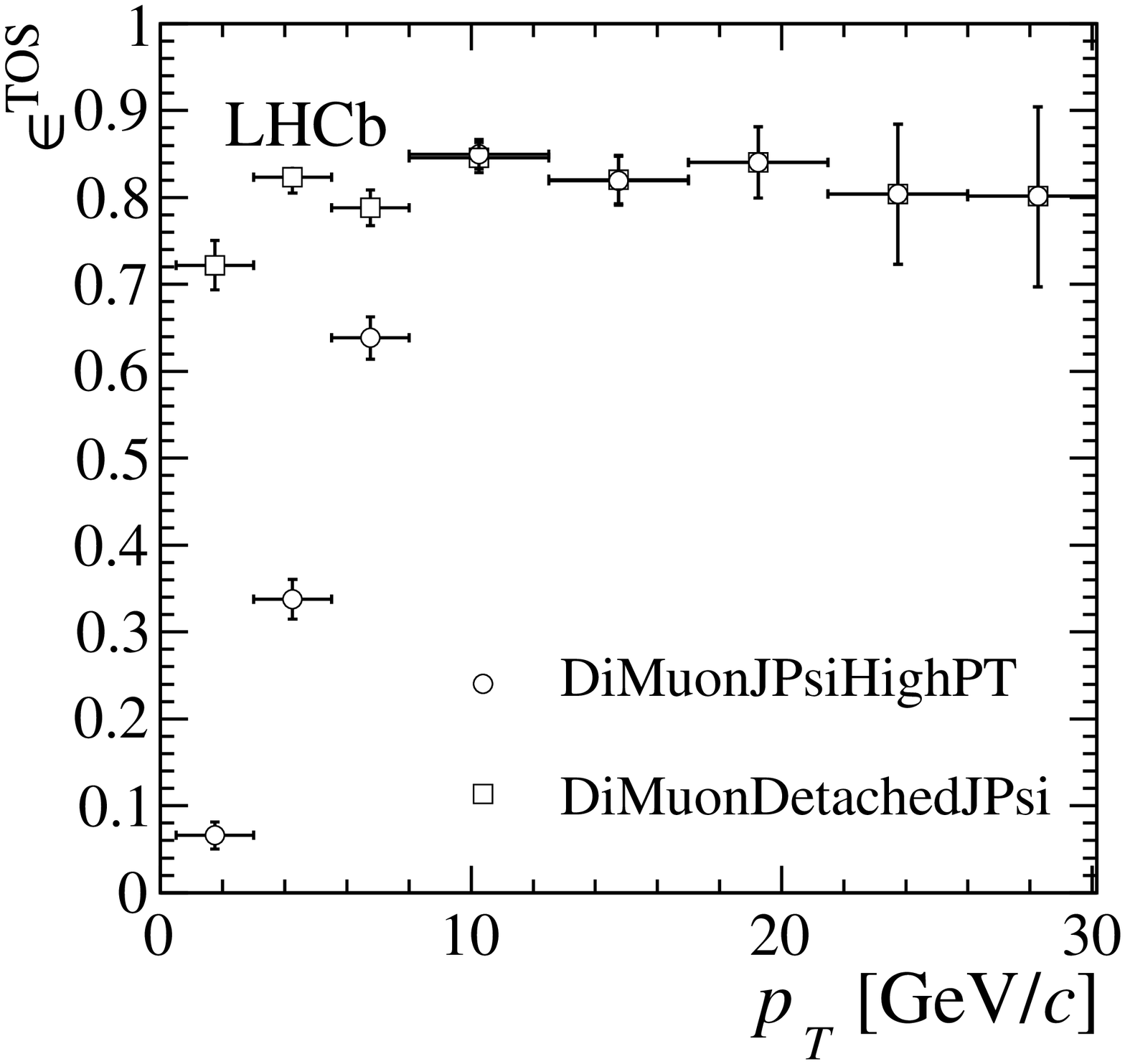,width=0.5\textwidth}}
\subfigure{\epsfig{figure=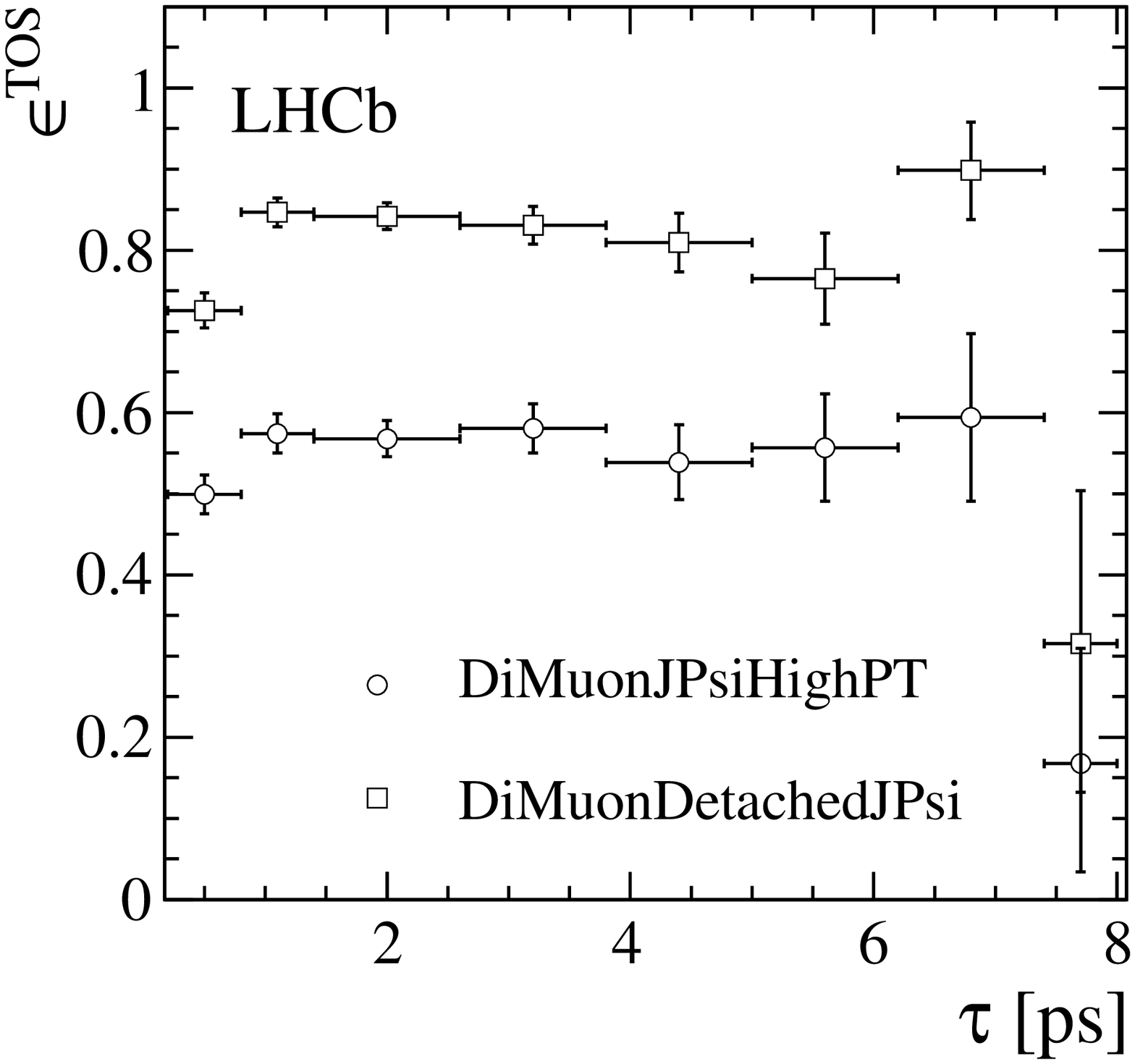,width=0.5\textwidth}}}
\caption{Efficiencies $\epsilon^{\rm TOS}$ of {\tt Hlt2DiMuonJPsiHighPT} and {\tt Hlt2DiMuonDetachedJPsi} 
for \BuToJPsiK as a function of \pt and $\tau$ of the $B^+$.}
  \label{fig:hlt2dimuon}
\end{figure}

The names and corresponding cuts of the detached decay selections are given in 
Table~\ref{tab:hlt2dimudetached}.
\begin{table*}[ht]
\caption{HLT2 lines based on two identified muons.
}
\label{tab:hlt2dimudetached}
\begin{center}
\begin{tabular}{l|r |r |r }
{\tt Hlt2DiMuon} &{\tt Detached}&{\tt DetachedHeavy}&{\tt DetachedJPsi}\\\hline
Track $\chi^2/\rm ndf$&$<5$&$<5$&$<5$\\
Track IP$\chi^2$&$>9$&-&-\\
Mass [\gevcc]&$>1$&$>2.95$&M$_{J/\psi}\pm 0.12$\\
${\rm FD} \chi^2$&$>$49&$>$25&$>$9\\
$\chi_{\rm vertex}^2$&$<25$&$<25$&$<25$\\
$\pt^{\mu\mu}$ [\gevc]& $>1.5$&-&-\\
\hline
Rate [Hz]&70&75&35\\
\end{tabular}
\end{center}
\end{table*}
{\tt Hlt2DiMuonDetached} is the main trigger for low mass muon pairs. 
{\tt Hlt2DiMuonDetachedHeavy} is an analogous trigger line for \jpsi and higher mass muon pairs, with
relaxed lifetime selection criteria.
{\tt Hlt2DiMuonDetachedJPsi} enhances the efficiency for \jpsi by reducing the flight distance
requirement for \jpsi candidates even further.
Figure~\ref{fig:hlt2dimuon} compares the performance of two representative {\tt HLT2DiMuon} lines: {\tt Hlt2DiMuonJPsiHighPT} and {\tt Hlt2DiMuonDetachedJPsi} in \BuToJPsiK
decays.  
{\tt Hlt2DiMuonJPsiHighPT} avoids by design a bias in the proper lifetime, at the price of losing efficiency at low $\pt(\jpsi)$.
The detached lines allow the selection of decays with a significant flight distance
with high efficiency even at low $\pt(\jpsi)$, but their efficiency is reduced at small lifetimes. 

\sloppy
There are nine BBDT topological lines: {\tt Hlt2Topo$n$Body}, {\tt Hlt2TopoMu$n$Body} and 
{\tt Hlt2TopoE$n$Body}, where $n$=2,3,4 for the
multiplicities considered. {\tt TopoMu} ({\tt TopoE}) require at least one of the decay particles to have been identified
as a muon (electron). Each line returns an output of the BBDT between 0 and 1. {\tt Hlt2Topo$n$Body} lines accept 
events with a combined rate of 930\,Hz with a cut on the BBDT output at 0.4, 0.4 and 0.3 for the 2, 3 and 4 body lines respectively.
While the {\tt TopoMu} and {\tt TopoE} lines are based on the same BBDT, the extra requirement of either a muon or
electron allows the cut on the BBDT output to be reduced to 0.1 for all six lines, which results in rates of
290 and 260\,Hz for {\tt TopoMu} and {\tt TopoE} respectively.

\begin{figure}[htb]
 \centering
\mbox{\subfigure{\epsfig{figure=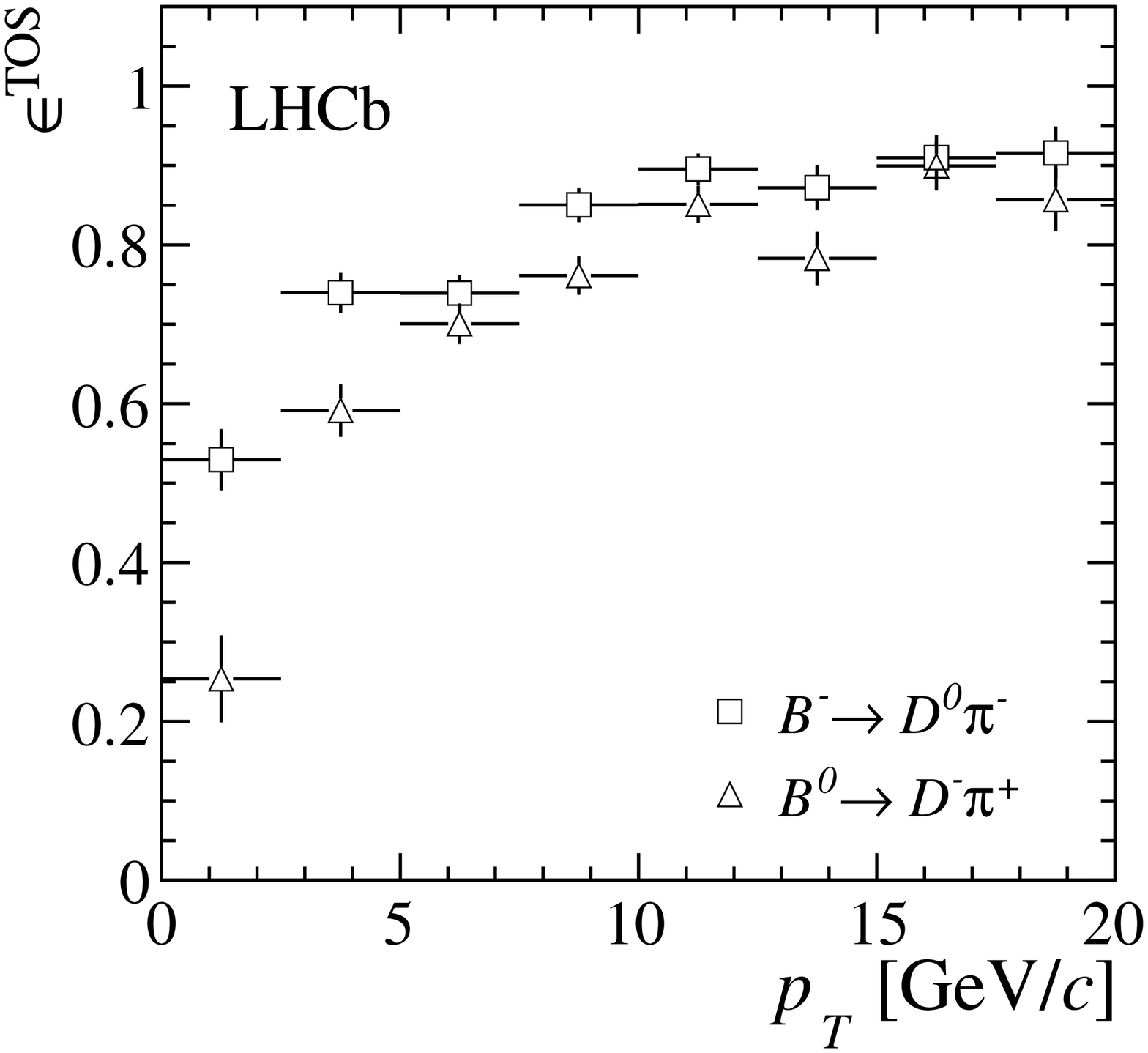,width=0.5\textwidth}}
\subfigure{\epsfig{figure=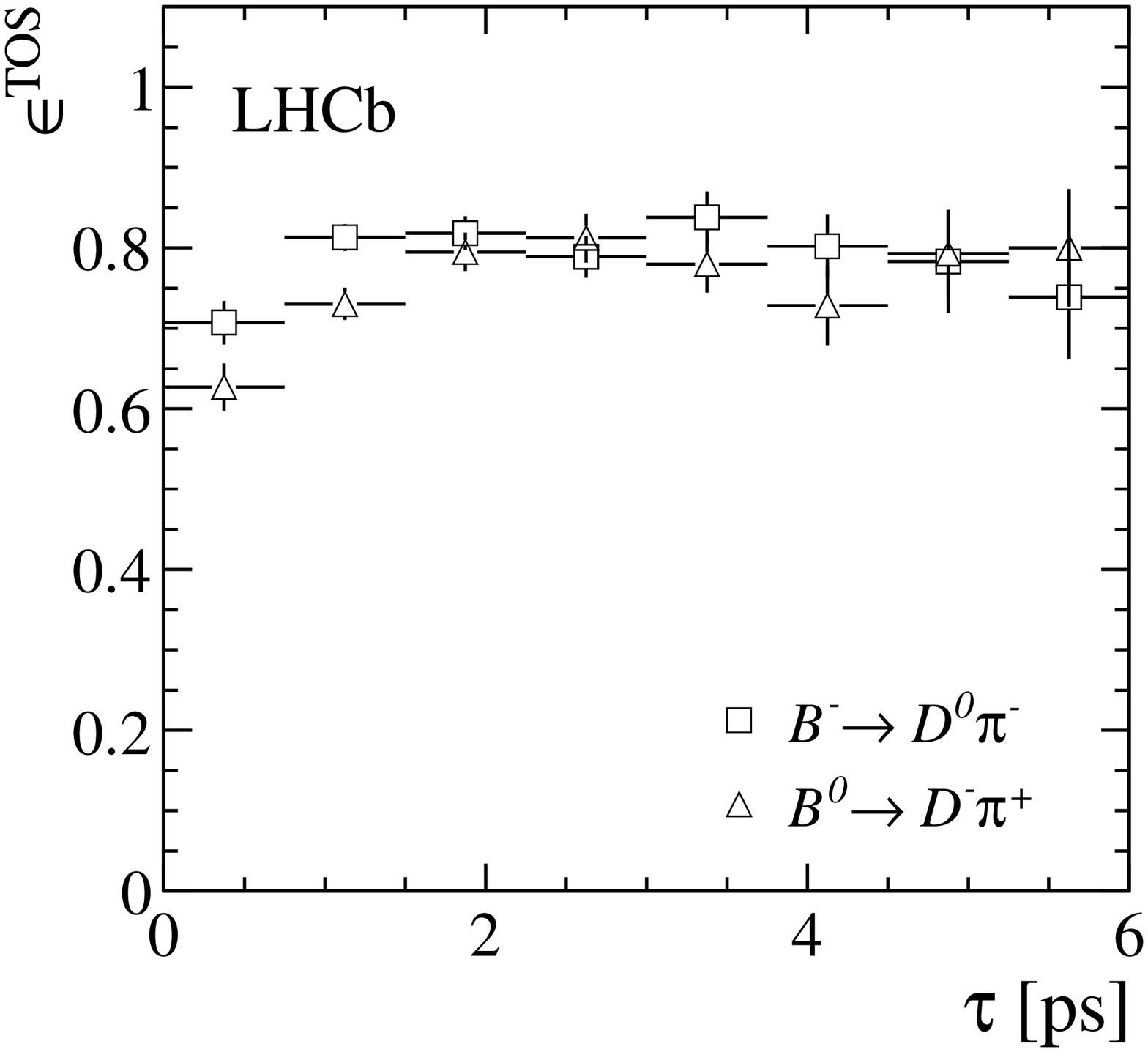,width=0.5\textwidth}}}
 \caption{
Efficiency $\epsilon^{\rm TOS}$ if at least one of the lines {\tt Hlt2Topo$n$Body}, with $n=2,3$, selected the event 
for \BuToDpi and one of the lines with $n=2,3,4$ for \BdToDpi as a 
function of \pt and  $\tau$ of the \B-meson.
The efficiency is measured relative to events that are TOS in {\tt Hlt1TrackAllL0}.
 }
 \label{fig:topo-eff-had}
\end{figure}

\begin{figure}[htb]
 \centering
\mbox{\subfigure{\epsfig{figure=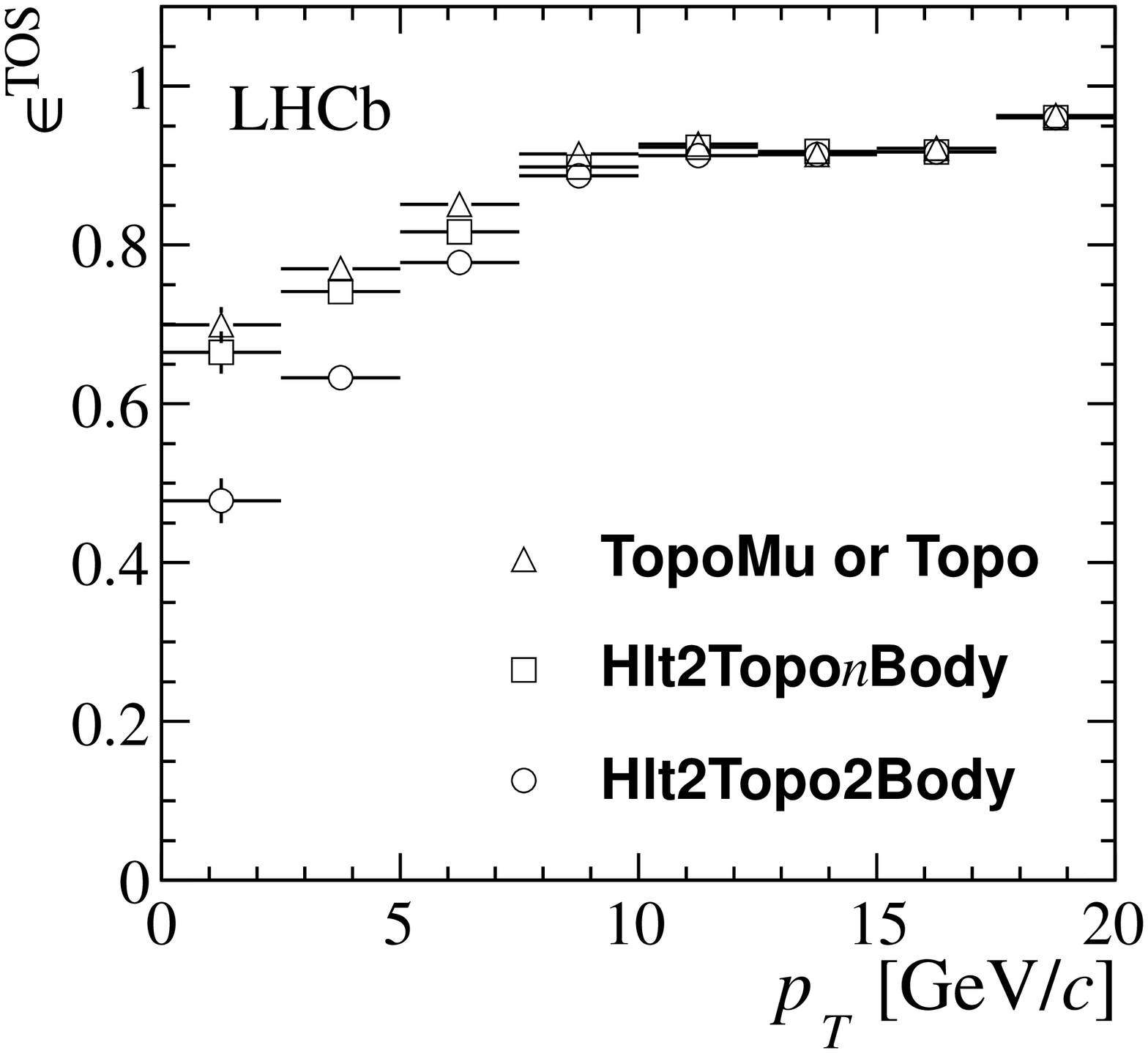,width=0.5\textwidth}}
\subfigure{\epsfig{figure=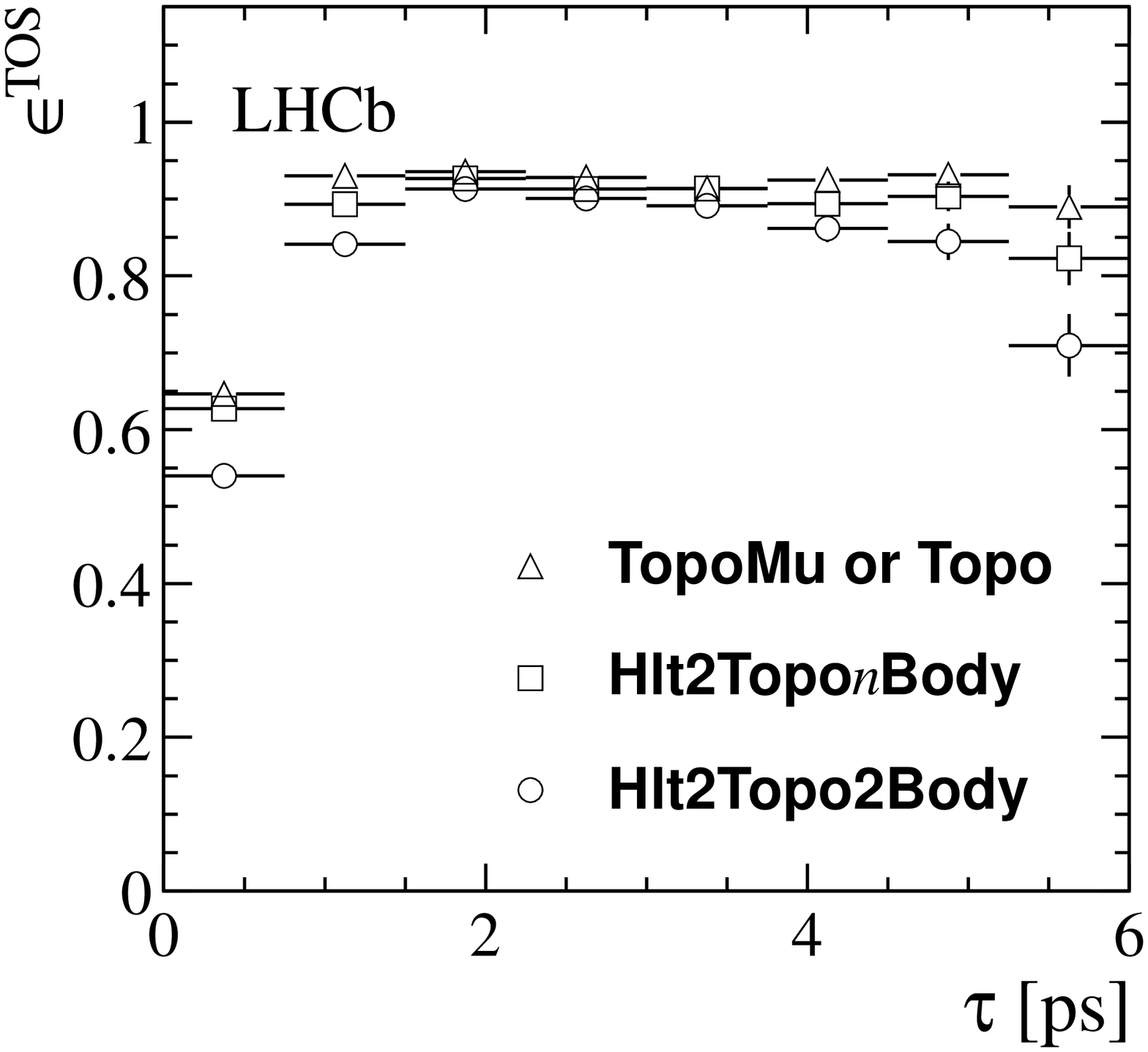,width=0.5\textwidth}}}
\caption{
Efficiency $\epsilon^{\rm TOS}$ if at least one of the lines {\tt Hlt2Topo$n$Body} or  {\tt Hlt2TopoMu$n$Body}, with $n=2,3$, selected events for 
\BuToJPsiK, 
as a function of \pt and $\tau$ of the \B-meson.
Also shown is $\epsilon^{\rm TOS}$ if the line {\tt Hlt2Topo$n$Body}, with $n=2,3$, 
selected the events.
{\tt Hlt2Topo2Body} shows the inclusive performance of the topological lines.
The efficiency is measured relative to events that are TOS in either {\tt Hlt1TrackAllL0} 
or {\tt Hlt1TrackMuon}.}
 \label{fig:topo-eff-mu}
\end{figure}

\sloppy
The performance of the topological lines is given in Fig.~\ref{fig:topo-eff-had} for fully hadronic
\B-decays and Fig.~\ref{fig:topo-eff-mu} for \BuToJPsiK decays. 
Figure~\ref{fig:topo-eff-mu} also shows the complementarity of {\tt Hlt2Topo$n$Body} and {\tt
Hlt2TopoMu$n$Body}; the efficiency increases if either of these lines has selected the signal event.
The inclusive performance of the topological lines is demonstrated in Fig.~\ref{fig:topo-eff-mu} by giving the performance of {\tt Hlt2Topo2Body} alone. This line
requires only two of the three decay tracks of \BuToJPsiK to have been reconstructed and selected. Adding {\tt Hlt2Topo3Body}
mainly recovers efficiency at low \pt compared to  
the {\tt Hlt2Topo2Body} line alone.

\begin{figure}[htb]
 \centering
\mbox{\subfigure{\epsfig{figure=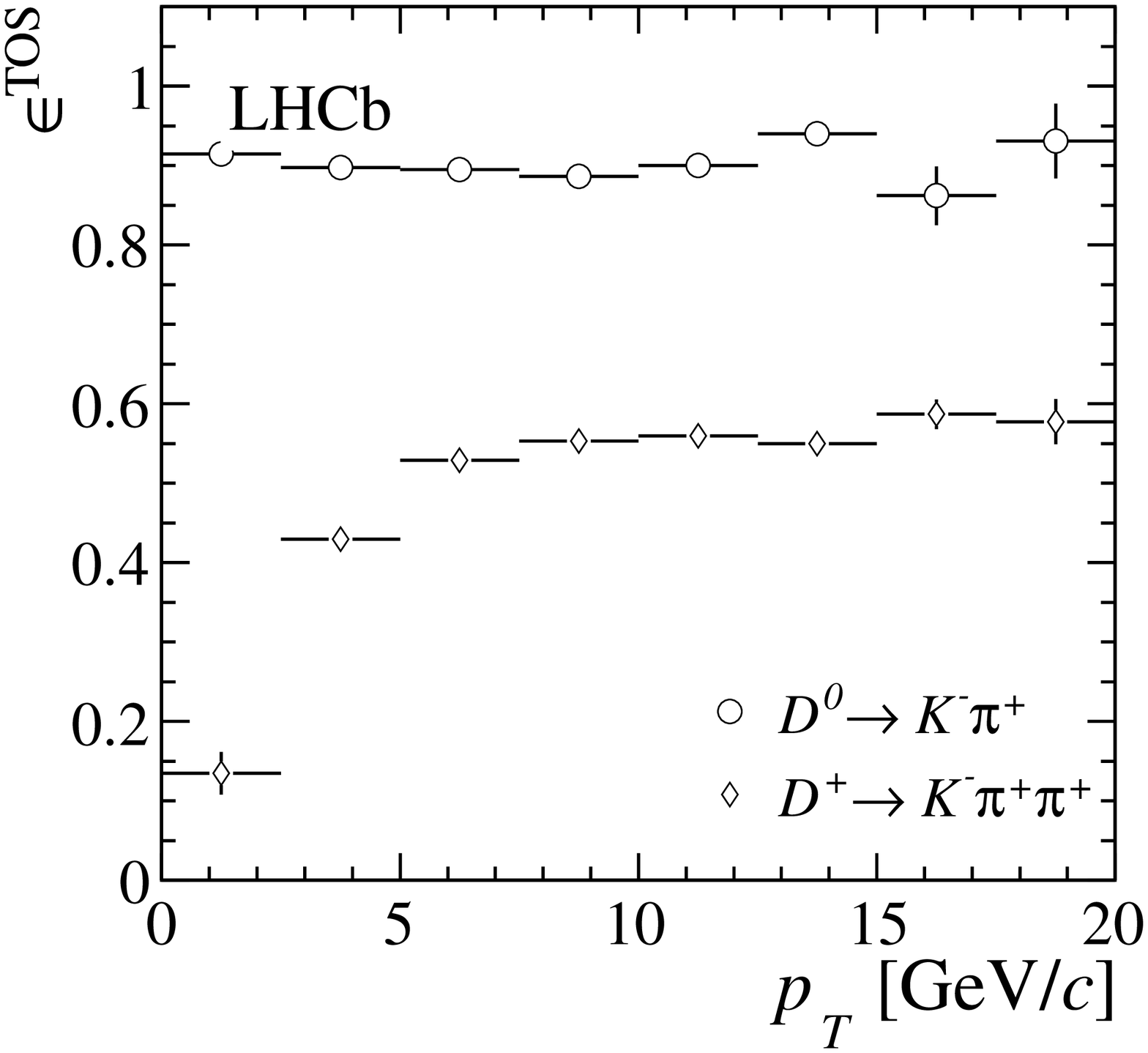,width=0.5\textwidth}}
\subfigure{\epsfig{figure=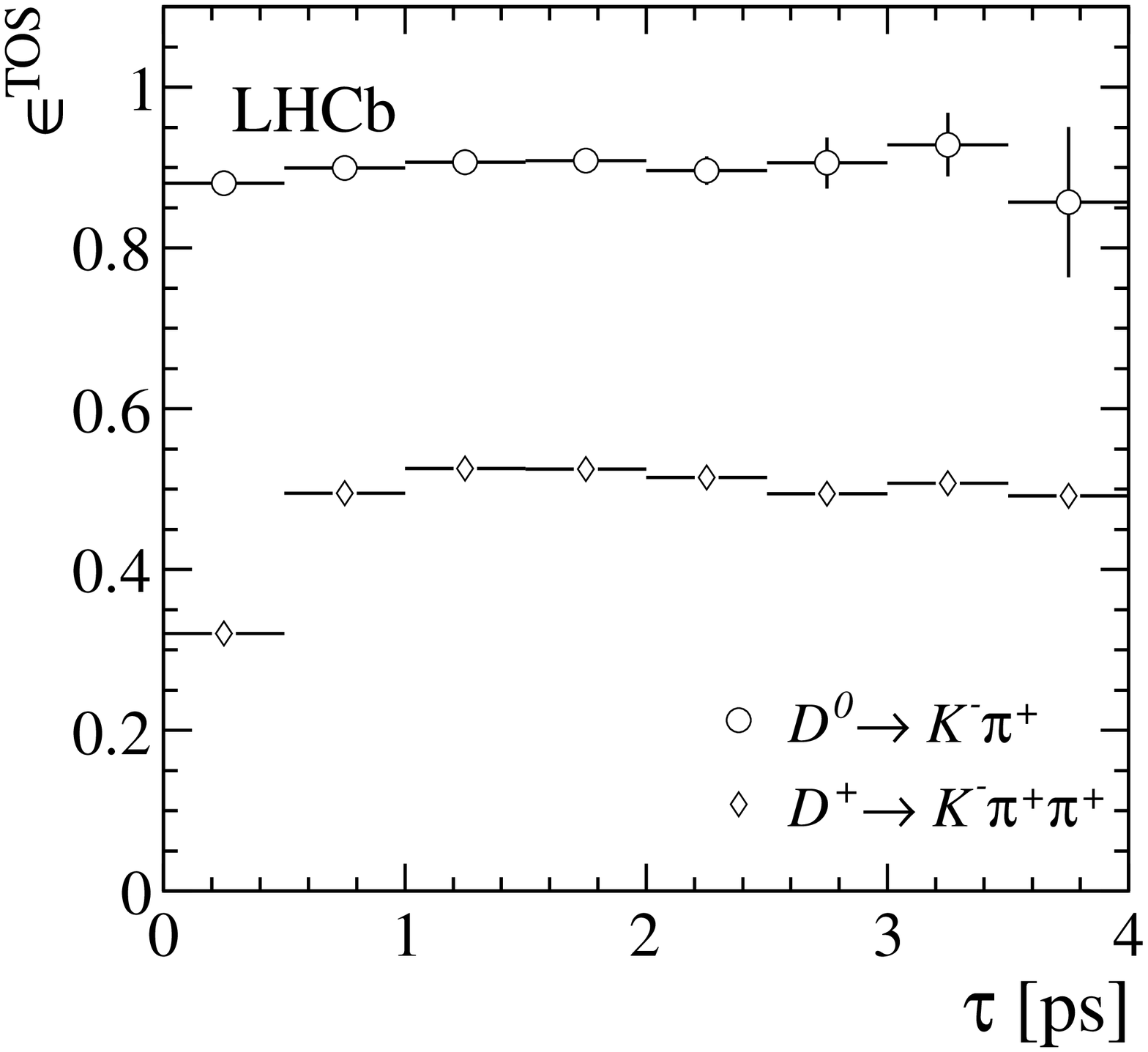,width=0.5\textwidth}}}
 \caption{
Efficiency $\epsilon^{\rm TOS}$ of the lines {\tt Hlt2CharmHadD2HHH} and {\tt Hlt2CharmHadD02HH$\_$D02KPi}
for $D^+\rightarrow K^-\pi^+\pi^+$ and $D^0\rightarrow K^-\pi^+$ respectively as a
function of \pt and  $\tau$ of the \D-meson.
The efficiency is measured relative to events that are TOS in {\tt Hlt1TrackAllL0}.}
 \label{fig:hlt2-d-eff-had}
\end{figure}

\sloppy
Table~\ref{t:hlt2d2hh-cuts} lists the cuts applied in the two HLT2 exclusive lines  {\tt Hlt2CharmHadD02HH$\_$D02KPi} 
and {\tt Hlt2CharmHadD2HHH}. The off-line selections of $D^0\rightarrow K^-\pi^+$ are only slightly tighter than
the cuts applied in HLT2, resulting in an almost maximum efficiency of the {\tt Hlt2CharmHadD02HH$\_$D02KPi} line for
this channel, as shown in Fig.~\ref{fig:hlt2-d-eff-had}. This figure also shows the
performance of  {\tt Hlt2CharmHadD2HHH} for $D^+\rightarrow K^-\pi^+\pi^+$. Here HLT2 
loses efficiency due to the necessity of first having to apply hard cuts 
to two of the three decay products before allowing an extra reconstruction step for low \pt tracks, as described in Section ~\ref{sec:exclusive}.
\begin{table}
  \begin{center}  
  \caption{HLT2 selection cuts applied for the exclusive lines {\tt Hlt2CharmHadD02HH$\_$D02KPi} and 
{\tt Hlt2CharmHadD2HHH}. The 2-track cuts refer to a candidate constructed of two tracks, 
and $m_{\rm corr}$ is defined in equation~2.
The angle $\alpha$ is the angle between the momentum of the \D and the vector connecting the PV with the \D vertex.
Some selections require that at least one or two tracks pass a cut, indicated with "$\geqq $". }
  \label{t:hlt2d2hh-cuts} 
  \begin{tabular}{ l|c|c }
    Variable & {\tt Hlt2CharmHadD02HH$\_$D02KPi} & {\tt Hlt2CharmHadD2HHH}\\\hline
    $\chi_{\rm track}^2/\rm ndf$& $<3$&  $<$3 \\
    Track \pt  [\mevc]              &$>$  800& $>$250 \\
    Track \ptot    [\mevc]              &$>5000$& $>$2000  \\
    Track $\Sigma$\pt         [\mevc]              &-& $>$             2500\\
    $\geqq 1$ track \pt     [\mevc]           &$>1500$& -   \\
    2-track mass [\mevcc] &-&$<$ 2100\\
    2-track $m_{\rm corr}$ [\mevcc] &-&$<$3500\\
    2-track IP$\chi^2$&-&$>$40\\
    $\geqq 2$ tracks \pt    [\mevc]            &- & $>$500   \\
    $\geqq 2$ tracks \ptot  [\mevc]              &- & $>$5000   \\
    Track IP$\chi^2$              &  $>9$ & $>5$   \\
    $\geqq 2$ tracks IP$\chi^2$              &  - & $>10$   \\
    2-track DOCA [mm]                    &-& $<$0.1  \\
    $\chi_{\rm vertex}^2/\rm ndf$& $<10$  &$<20$\\
    FD$\chi^2$                   &$>40$& $>$ 150\\
    $$D IP\chisq                  &-& $<$              12\\
    $D$ cos($\alpha$) & $>0.99985$&- \\
    $D$ \pt           [\mevc]              &$>2000$& $>$             1000\\
    $D$ mass interval [\mevcc] & 1815-1915 &1800-2040\\\hline
Rate [Hz]& 260 & 390\\
  \end{tabular}
  \end{center}
\end{table} 
\section{Summary}
\label{sec:conclusions}
\fussy
The LHCb trigger is designed to select charm and beauty hadrons in a large range of decay modes, and
permits the 
measurement of its efficiency directly from data. In 2011 the trigger has been tuned to cope with $pp$ interactions at
$\sqrt{s}=7$\,TeV, with 1296 colliding bunches in LHCb and an average number of visible $pp$ interactions per bunch crossing of 1.4.
This corresponds to a bunch crossing rate with at least one visible $pp$ interaction of $\sim 11$\,MHz. 

L0 reduces this rate to 870 kHz by applying \pt cuts on muons and $E_T$ cuts on clusters in
the calorimeters. HLT1 performs a partial reconstruction of tracks and performs muon identification. It employs
a combination of cuts on \pt, invariant mass and IP to reduce the rate to around 43\,kHz. HLT2 reconstructs all
tracks in the event with \pt$>$500 \mevc. It selects candidates based on lepton identification, lifetime information
and invariant mass. Its output rate is 3\,kHz, consisting of 50\,$\%$ inclusive hadronic triggers,
25\,$\%$ triggers on leptons and the remaining rate from exclusive triggers, mainly on charmed
hadrons. The efficiencies for the major trigger lines are presented for
representative decay modes as a function of \pt and lifetime of $c$ and $b$-hadrons. 

The successful exploitation of the LHC as a beauty factory relies crucially on the ability to trigger on heavy flavour decays in a hadronic 
environment.
To achieve this, the trigger is designed to be able to determine the impact
parameter of tracks at a high rate, and to measure the momentum of those tracks with sufficiently
large impact
parameter, or to identify them as muon candidates. 
The trigger managed to adapt to the larger pile-up conditions imposed by the machine delivering only 1296 instead of the planned 2622 colliding 
bunches in the LHC. The trigger performance and the fact that its efficiency can be evaluated in a
data-driven way, in combination with the excellent 
performance of the sub-detectors, allowed LHCb to publish more than 40 papers based on the data collected in 2011.

LHCb is preparing to upgrade the detector~\cite{upgrade} in 2018. It will feature a fully software based trigger that will allow it to explore its physics goals 
at even larger luminosities.

\acknowledgments
We express our gratitude to our colleagues in the CERN
accelerator departments for the excellent performance of the LHC. We
thank the technical and administrative staff at the LHCb
institutes. We acknowledge support from CERN and from the national
agencies: CAPES, CNPq, FAPERJ and FINEP (Brazil); NSFC (China);
CNRS/IN2P3 and Region Auvergne (France); BMBF, DFG, HGF and MPG
(Germany); SFI (Ireland); INFN (Italy); FOM and NWO (The Netherlands);
SCSR (Poland); ANCS/IFA (Romania); MinES, Rosatom, RFBR and NRC
``Kurchatov Institute'' (Russia); MinECo, XuntaGal and GENCAT (Spain);
SNSF and SER (Switzerland); NAS Ukraine (Ukraine); STFC (United
Kingdom); NSF (USA). We also acknowledge the support received from the
ERC under FP7. The Tier1 computing centres are supported by IN2P3
(France), KIT and BMBF (Germany), INFN (Italy), NWO and SURF (The
Netherlands), PIC (Spain), GridPP (United Kingdom). We are thankful
for the computing resources put at our disposal by Yandex LLC
(Russia), as well as to the communities behind the multiple open
source software packages that we depend on.
A special acknowledgement goes to all our LHCb collaborators who over the years
have contributed to obtain the results presented in this paper.

\end{document}